 \documentclass[preprint, showpacs, superscriptaddress, citeautoscript, aps, pre, longbibliography]{revtex4-1}
\usepackage{graphicx}
\usepackage{color}
\usepackage{amssymb}
\usepackage{amsmath}
\usepackage{subcaption}
\usepackage{verbatim}
\usepackage{hyperref}
\usepackage{siunitx}
\usepackage{braket}
\usepackage{multirow}
\usepackage{xspace}
\usepackage{listings}
\usepackage{tikz} 

\newcommand{\etal}{et al.\@\xspace}
\DeclareSIUnit\angstrom{\text{Å}}

\makeatletter
\newcommand*\fsize{\dimexpr\f@size pt\relax}
\makeatother

\hypersetup{
    colorlinks=true, 
    linktoc=all,     
    linkcolor=blue,  
    citecolor=blue,
}

\begin{document}

\title{Optimal Invariant Bases for Atomistic Machine Learning}
\author{Alice E. A. Allen}
\affiliation{Theoretical Division, Los Alamos National Laboratory, Los Alamos, New Mexico 87546} 
\affiliation{Center for Nonlinear Studies, Los Alamos National Laboratory, Los Alamos, New Mexico 87545}
\affiliation{Max Planck Institute for Polymer Research, Ackermannweg 10, 55128 Mainz, Germany}

\author{Emily Shinkle}
\affiliation{Computer, Computational, and Statistical Sciences Division, Los Alamos National Laboratory, Los Alamos, NM 87545, USA}

\author{Roxana Bujack}
\affiliation{Computer, Computational, and Statistical Sciences Division, Los Alamos National Laboratory, Los Alamos, NM 87545, USA}

\author{Nicholas Lubbers}
\email{nlubbers@lanl.gov}
\affiliation{Computer, Computational, and Statistical Sciences Division, Los Alamos National Laboratory, Los Alamos, NM 87545, USA}

\date{\today}

\begin{abstract}
The representation of atomic configurations for  machine learning models has led to the development of numerous descriptors, often to describe the local environment of atoms. However, many of these representations are incomplete and/or functionally dependent. Incomplete descriptor sets are unable to represent all meaningful changes in the atomic environment. Complete constructions of atomic environment descriptors, on the other hand, often suffer from a high degree of functional dependence, where some descriptors can be written as functions of the others. These redundant descriptors do not provide additional power to discriminate between different atomic environments and increase the computational burden. By employing techniques from the pattern recognition literature to  existing atomistic representations, we remove  descriptors that are functions of other descriptors to produce the smallest possible set that satisfies completeness. We apply this in two ways: first we refine an existing description, the Atomistic Cluster Expansion. We show that this yields a more efficient subset of descriptors. Second, we augment an incomplete construction based on a scalar neural network, yielding a new message-passing network architecture that can recognize up to 5-body patterns in each neuron by taking advantage of an optimal set of Cartesian tensor invariants. This architecture shows strong accuracy on state-of-the-art benchmarks while retaining low computational cost. Our results not only yield improved models, but point the way to classes of invariant bases that minimize cost while maximizing expressivity for a host of applications.
\end{abstract}
\maketitle

\section{Introduction}
\label{sec:intro}

Over the past decade, machine learning has emerged as a powerful tool for modeling atomistic systems~\cite{Botu2017MachineOutlook,Deringer2019Machine,extendingbeyondproperties,Unke2021Machine,kulichenko2021rise,Behler2021four,Musil2021Physics}.
An essential component of machine learning (ML) for atomistic systems is building efficient and effective representations of local regions in a system---often through an atom-centered perspective. A significant milestone in the development of machine learning models for atomistic systems was recognizing the importance of fundamental symmetries. 
Early approaches to machine learning in atomistic systems did not always consider symmetry~\cite{No1997, Chmiela2017}, which  reduced data efficiency and performance~\cite{Chmiela2018}. Subsequently, the importance of symmetry was recognized~\cite{Behler2007, Braams2009-wi}, which spurred the development of several successful ML approaches~\cite{GAPoriginal,Bartok2013OnEnvironments,Thompson2015SpectralPotentials,Shapeev2015MomentPotentials,Smith2017ANI-1:Costb,Schutt2017Quantum-ChemicalNetworks,Schutt2018SchNetMaterials,Lubbers2018,Unke2019PhysNet:Charges,Zubatyuk2019AccurateNetwork,linsey2017chimes,allen2021atomic} based on the characterization of the atomic environment in terms of scalar features.
It was realized by Pozdnyakov \etal that many of these representations are incomplete, that is, that they cannot distinguish all atomic environments~\cite{Pozdnyakov2020}.
In search of further expressive capacity, tensors, which transform under rotational symmetry in a well-behaved way, have been used~\cite{thomas2018tensorfieldnetworksrotation,allegro,Gasteiger2020Directional,wang2024equivariant,e3equivariant,Shapeev2015MomentPotentials,allegro,Gasteiger2020Directional,Chigaev2023,luo2024nabling,Nigam2020nice,Kovacs2021,Lysogorskiy2021,AtomicClusterExpansion,batatia2022design,batatia2022mace,Frank2024Euclidean,Kabylda2025molecular}, mostly built using either the tensor product operation~\cite{thomas2018tensorfieldnetworksrotation} to define higher-order tensor representations in terms of products of lower-order representations, or using a projection of a many-body density onto a spherical basis~\cite{Bartok2013OnEnvironments,Shapeev2015MomentPotentials,AtomicClusterExpansion}. 

The atomic cluster expansion (ACE)~\cite{AtomicClusterExpansion} has proven to be an important milestone, defining a complete basis set for functions of a local atomic environment, for all tensor orders, by exhaustive enumeration~\cite{AtomicClusterExpansion}. In some literature discussing ACE, a different, stronger notion of completeness is discussed---the capability to describe functions of an atomic basis set as a linear expansion. In mathematical literature this notion may be referred to by describing a set of descriptors as \emph{dense} in the space of atomic environment functions; a linear ACE model is, loosely speaking, a universal approximator for atomic environment functions. However, because it contains a number of descriptors that grows combinatorically with the series parameters, the ACE construction is expensive to use at high order, thus motivating the search for reduced (that is, more effective) ways to represent environments within the ACE framework~\cite{Nigam2020,batatia2022mace,Dusson2022,Cheng2024,bochkarev2024graph}. 

 On the other hand, knowing that an atom has $k$ neighbors, there are only $3k-3$ degrees of freedom needed to specify the environment up to rotational symmetry (ignoring, for the moment, atomic species). This suggests that it may be possible to construct a set of functions that can serve as coordinates to identify any atomic environment up to rotational symmetry that might be far smaller and computationally tractable. Indeed, a recent study revealed that by using a triple-centered instead of an atom-centered representation, a complete set of descriptors can be resolved that has  $O(k)$ elements~\cite{nigam2024completeness}. Universal approximation theorems~\cite{hornik1989multilayer} imply that if we can build neurons that can recognize arbitrary locations in the input space, a multi-layer neural network can reproduce arbitrary functions on that space to arbitrary levels of accuracy. In other words, by creating a complete set of symmetry-invariant descriptors, the descriptors can be fed through standard machine learning methods with the knowledge that given enough training data, the model can represent the complex functions that describe atomistic systems such as molecules and materials.

The mathematics of building a neuron that can recognize arbitrary locations can be found in the literature for pattern recognition, as a single neuron can be thought of as a pattern detector. Pattern detection under rotational symmetries is well-studied~\cite{Hu62, DN77, Flu00, LH09, FSZ16, bujack2022systematic}. A plethora of invariant functions has been proposed to characterize patterns in a rotation-invariant way. As such, a fruitful line of attack is to ask how to prune these functions by eliminating those that do not provide additional power for discriminating different signals. One wants to examine \emph{functional dependence}---whether an invariant can be written as a function of other invariants.
Jacobi defined the classic approach to this type of problem by examining the determinant of a matrix of partial derivatives over a number of variables---the key result being that functional dependence implies the linear dependence of derivatives~\cite{jacobi1841determinantibus}. More recently, Langbein \etal provided a useful algorithm for applying this approach to a large number of functions, yielding a set of invariants which are functionally independent---a set of functions for which no element can be written as a function of other elements~\cite{LH09}. Important gaps in applying this approach to the detection of spatial patterns were identified and addressed by Bujack \etal, who developed the notion of a \textit{minimal flexible set}---a minimal set of invariants which is complete for any admissible pattern~\cite{bujack2017moment, bujack2022systematic}. Generalizations of this approach to spherical functions and irreducible representations have recently been developed, in a companion paper to our work, by Bujack \etal~\cite{bujack2025flexible}

In this paper, we apply these pattern detection principles in two ways to machine learning interatomic potentials (MLIPs), one of the dominant types of atomistic ML models. First, to investigate the minimal set of functionally-independent ACE descriptors, we have implemented a form of the Langbein algorithm for ACE\@. By analyzing the Jacobian of a set of descriptors, we can arrive at a subset of ACE which we then apply to six single-element materials introduced in Ref~\citenum{Zuo2020}. For a given basis-set size, the functionally-independent subset consistently offers superior performance in force prediction across these materials.  This could be of particular interest for non-linear forms of ACE, as placing a complete set of descriptors into a non-linear function enables the reproduction of arbitrary functions. Second, we use a flexible set of Cartesian tensor invariants to build neurons that can individually recognize any pattern without a combinatorial explosion in the tensor expansion orders. These neurons can be used to define a complete point-cloud neural network for scalars, which we build as an extension to HIP-NN~\cite{Lubbers2018};  we refer to this high-order-polynomial version of the architecture as HIP-HOP-NN\@.  Each neuron in HIP-HOP-NN can recognize arbitrary 5-body patterns in atomic-environment features. We test the architecture on several benchmark datasets and conclude that it has clear benefits over previous HIP-NN models, and compares well to other state-of-the-art architectures, providing strong accuracy at reduced computational expense. Overall, our results show that feature sets which are optimized using pattern recognition techniques can improve the cost-accuracy Pareto front of atomistic ML models, and the technique can be applied to virtually any approach which uses explicit feature sets.

\section{Methods}
\label{sec:Methods}

\subsection{Overall Strategies}

The aim of this work is to apply strategies of \emph{optimal} pattern recognition to problems for atomistic machine learning.  We will begin in section~\ref{sec:pattern_recognition} by explaining these strategies in the abstract, applied to a generic set of invariant features, to assemble reduced sets of invariant features. These reduced sets are independent in a \emph{nonlinear} sense, using a local test. We also explain the need for constructing \emph{flexible} sets which are functionally independent in a global sense. In section \ref{sec:ace_formulation} we review how spherical tensors are used to build potentials in linear ACE, and in section \ref{sec:ace_langbein} we then explain how to apply the aforementioned filtering strategies to reduce the basis set of the ACE which uses these spherical tensors. In section \ref{sec:cartesian_graphs}
we review an alternative to the spherical strategy for enumerating a set of invariants, instead using the equivalent Cartesian tensor formulation. Using Cartesian tensors, the simple invariant expressions correspond to multigraphs. Passing these multigraphs through the flexible-basis search methods yields a relatively small set of tensors that extracts all functionally \emph{independent} information from a function on the sphere. This set does not grow combinatorically in the expansion order, instead growing only quadratically, which is asymptotically identical to the number of degrees of freedom in the tensors themselves~\cite{bujack2025flexible}. Finally, in section \ref{sec:hip-hop-methods} we detail how we have incorporated this set of invariants into a neural network architecture based on HIP-NN, which we call HIP-HOP-NN.

\subsection{Pattern recognition based on the method of Langbein and Flexible Basis Sets}
\label{sec:pattern_recognition}
Moment invariants~\cite{Reiss1991}, which themselves have been used for MLIPs~\cite{Shapeev2015MomentPotentials,Uhrin2021}, have been studied extensively in the field of pattern recognition. 3D geometric moments are projections of a scalar function $f:\mathbb R^3\to\mathbb R$ to the monomial basis. The moments  of each order $\ell$ form a moment tensor 
\begin{equation}\label{momentTensor} \begin{aligned}
{}{^\ell}M_{\alpha_0...\alpha_{l-1}}&=\int r_{\alpha_0}...r_{\alpha_{\ell-1}}f(\mathbf{r})d^3 \mathbf{r}
\end{aligned}\end{equation}
of rank $\ell$~\cite{LD89}, where the $\alpha_i\in\{1,2,3\}$ indicate one coordinate of $\mathbf{r}\in\mathbb R^3$. Products and contractions of tensors are tensors, and zeroth-rank tensors are invariant to rotations. This allows us to generate sets of invariant descriptors by multiplying and contracting moment tensors in all combinations. A key drawback is that this produces a vast number of invariants and most are redundant in some way.

Optimal bases of moment invariants in pattern recognition are \emph{complete, functionally independent, and flexible}~\cite{bujack2017moment}. Complete means that any two functions that differ by more than a rotation have different descriptors. Functionally independent means that no descriptor can be written as a function of the others. Flexible means that the basis never becomes degenerate. These three criteria guarantee that the basis has full discriminative power and the smallest possible size.

The set of all possible total contractions of the moment tensors form a complete set of rotationally invariant descriptors, but the size of this set scales combinatorically with the maximum tensor order. The number of functionally-independent moment invariants is given by the number of moments minus the dimensionality associated with the symmetry operation, which is in this case the three parameters required to specify a rotation. A functionally-independent set can theoretically be identified using the Jacobi criterion~\cite{jacobi1841determinantibus}. 
The key is to recognize that if an invariant $I_n(m)$ as a function of the underlying moments $m$ is functionally dependent on other $I_0(m)$, \ldots, $I_{n-1}(m)$ invariants, then we can write an implicit function $f$ for $I_n(m)$:
\begin{equation}
    I_n(m) = f(I_1(m),\ldots,I_{n-1}(m)).
\end{equation}
As a consequence, the derivatives of the moments must be linearly dependent on each other, as
\begin{equation}
    \frac{\partial I_n}{\partial m} = \sum_{i=1}^{n-1} \frac{\partial f}{\partial I_i} \frac{\partial I_i}{\partial m}
\end{equation}
implies that the Jacobian of the moments, $\frac{\partial I_i}{\partial m}$, does not have full rank. By testing the Jacobian on candidate sets of moments, one can assemble a set of moments which is complete up to symmetry. 
Instead of analytically solving for the rank of the Jacobian, Langbein and Hagen developed an approximation algorithm that randomly assigns values to the moments and solves for the rank numerically~\cite{LH09}.

Combinatorically scaling sets of functionally-dependent tensor contractions has also been used in atomistic machine learning as Moment Tensor Potentials (MTP)~\cite{Shapeev2015MomentPotentials}. The moment tensors have the attractive property of extracting linearly-independent features from an atomic environment in a systematic way and forming a complete basis for regression. 

One of our contributions is recognizing that the strategy of Langbein and Hagen is applicable to filtering down any set of invariants used for featurizing a symmetric space, and in particular for atomic environments. The use of moment invariant methods from the 3D imaging community has been previously studied~\cite{Uhrin2021}, but in our work we show that the approach of Langbein and Hagen can be applied to a far wider range of descriptors than moment invariants alone, yielding more expressive ML models and improved results, and identified the necessity to invoke additional learning machinery, such as a neural network, over the moment invariants. 

Bujack \etal realized that the Langbein approach would not necessarily work when applied to certain patterns~\cite{bujack2022systematic}. This is because when symmetry causes a moment to vanish, a large number of moment invariants will vanish with it. As such, a set of invariants can be complete in a local region, but incomplete when considering the entire space of patterns. This necessitates building a \emph{flexible} basis: one which is complete across the space of possible patterns to detect. This issue becomes particularly important for applications in atomistic machine learning because the spherical nature of the functions causes naive bases of moment invariants to degenerate to incompleteness as discussed in Ref.~\citenum{bujack2025flexible}.
In this paper, we focus on demonstrating the applicability and effectiveness of this technique and not on the theoretical foundations. Explicit examples and the full derivation of a flexible basis from irreducible tensors are given in the companion paper~\cite{bujack2025flexible}.

\begin{figure*}
    \includegraphics[width=5in]{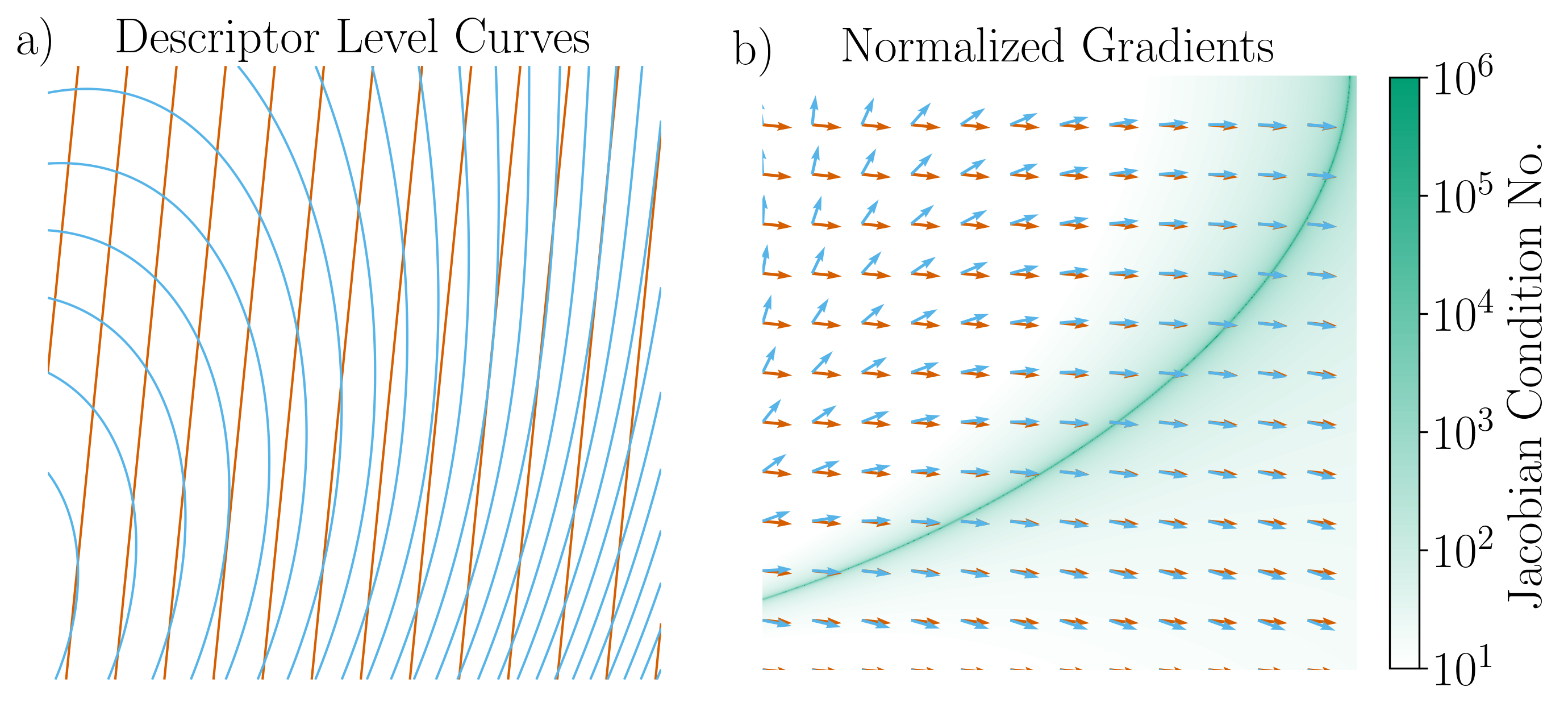}
    \caption{Abstract demonstration of the need for flexible basis sets. Left: a set of two descriptors visualized as iso-contours over the 2-D plane. Right: the gradients of the descriptors, as well as the condition number of the Jacobian. Because of the degeneracy of the Jacobian on the 1D manifold, each combination of descriptor values appears twice, and thus the corresponding functions cannot be differentiated by these two descriptors.}
    \label{fig:langbein_example}
\end{figure*}

The situation is depicted schematically in Fig~\ref{fig:langbein_example}. A set of descriptors is used to identify points on a manifold, in this case, the underlying plane. The $x$ and $y$ axes are two moments and the values of the functions as shown through iso-contours are two moment invariants. Locally at each point, the rank of the Jacobian of the descriptors corresponds to the dimension of the space spanned by the gradients of the descriptors at that point. Across most of the plane, these descriptors are functionally independent; the gradients are not parallel, so the rank of the Jacobian is two, and the Langbein algorithm identifies the descriptors as containing functionally-independent information. However, there is a sub-manifold on which these descriptors are actually functionally dependent on each other: on this sub-manifold, the level curves are parallel, the gradients are parallel, and the condition number of the Jacobian diverges because its rank is one. In other words, there is a line of critical points. Note that this does not require that the descriptors be linearly dependent on this manifold. As such, these two descriptors do not provide a flexible basis for characterizing the plane. One can also see this by tracking the intersection of the descriptors: for the level curves that intersect with the degenerate manifold, the descriptor level curves intersect at two points, and there is a set of points on one side of the degenerate manifold whose descriptors are identical to those found on the other side of the manifold. This means that the values of the invariants are identical at these two locations even though the values of the moments are different, illustrating that non-flexible bases can fail to differentiate functions even where they are not degenerate.

\subsection{The ACE Basis}

\label{sec:ace_formulation}

Here we briefly review the ACE basis for constructing features using spherical tensors. For a pedagogical discussion of spherical tensors for representing atomic environments, see Section~\ref{sec:functions_on_sphere}. We use the notation of Ref~\onlinecite{Kovacs2021} limited to one element type.
The atomic basis is defined as 
\begin{equation}
    \alpha_{nlm} = 
    \sum_{\substack{j}}
    \phi_{nlm}(\mathbf{r}_{ji})
\end{equation}
\begin{equation}
    \phi_{nlm}(\mathbf{r}) = R_{nl}(r)Y_l^m(\mathbf{\hat{r}}),
\end{equation}
where $R_{nl}$ is a radial basis function and $Y_l^m$ is a spherical harmonic function~\cite{AtomicClusterExpansion, Kovacs2021}. Because of the large number of indices associated with the atomic basis, they are collected into a single product-index $v = (n, l,m)$ in future formulas.
A permutation-invariant basis function is constructed by forming products of the atomic basis:
\begin{equation}
   A_{\mathbf{v}} = \prod_{t = 1}^\nu \alpha_{ v_t},\quad \mathbf{v} = (v_1, \dots, v_\nu).
   \label{eq:atomic_base}
\end{equation}
Rotational invariance is then achieved by the following formula, constituting an average of the multi-particle basis over all possible rotations:
\begin{align}
    B_{\mathbf{v}} & = 
    \int_{\hat R\in O(3)}
            \prod_{t = 1}^\nu A_{v_t}(\{\hat R \mathbf{r}_{ij}\})  \, d\hat R \label{eq:B-basis}\\
        &= \sum_{\mathbf{v}'} C_{\mathbf{v}\mathbf{v}'} A_{ \mathbf{v}'}, \label{eq:CG}
\end{align}
 where  the coefficients $C_{\mathbf{v}\mathbf{v}'}$ are sets of combined Clebsch-Gordan coupling coefficients that compute a scalar component of each element of $\mathbf{v}$. In a linear ACE model, the energy at site $i$ is then given by
 \begin{equation}
    E_i = \sum c_{\mathbf{v}} B_{\mathbf{v}} = \mathbf{c}\cdot\mathbf{B}.
    \label{eq:E_usingBs}
\end{equation}
A relevant aspect of ACE is that the basis $A_{\mathbf{v}}$, being formed of products, grows exponentially in the order $\nu$, which indexes the body-order of the features. This combinatoric growth propagates to the number of invariant features $\mathbf{B}$; for practical purposes one needs to define limitations on the maximum $n,l,\nu$.

For the computational aspects of this work, the ACEpotentials.jl code described in Ref.~\citenum{Witt2023} was used. The linear solver used for the Langbein algorithm is the rank-revealing QR factorization (RRQR) algorithm. This solver was chosen as it allows for fitting to singular matrices that occur at higher degrees. In this work, the weighting parameter between the radial and spherical harmonic components was set to 1.0, and the order was set to four. 

\subsection{A functionally-independent subset for ACE}
\label{sec:ace_langbein}
The combinatorically scaling number of features in ACE is not inherent to the problem but a result of many of the features containing redundant information in the form of functionally-dependent descriptors. For regression problems this redundant information can be useful for producing more expressive regression models. In this work, we adapt Langbein's algorithm to construct a functionally-independent subset of the ACE basis. This means that while maintaining its full discriminative power, we remove all descriptors that are functions of others to produce the smallest possible set that satisfies completeness. Our computation includes the following stages: 
\begin{itemize}
    \item  Construct a matrix of $\frac{\partial B_{\mathbf{v}}}{\partial \alpha_j}$ for one atom (from random neighbours). The components  $\frac{\partial B_{\mathbf{v}}}{\partial \alpha_j}$ are calculated numerically. 
    \item Initialize a candidate Jacobian matrix with columns equal to the number of atomic basis functions ($\alpha_{\mathbf{v}}$) and initially empty of rows.
    \item For each $B_{\mathbf{v}}$ until the utmost number of $B_{\mathbf{v}}$ terms is reached: compute whether appending the row vector of  partial derivatives of the  invariant will augment the rank of the candidate Jacobian matrix. If the rank is increased, append the partial derivatives of the invariant to the the candidate Jacobian matrix. From this, construct an index of $B_{\mathbf{v}}$ components that increase the rank of the candidate Jacobian matrix. 
    \item  Repeat this $N$ times (where $N$ is the number of data points sampled). Each repetition produces a new set of indices. A $B_{\mathbf{v}}$ member only needs to be identified as increasing the candidate Jacobian matrix rank for one data point to be included in the final subset. 
\end{itemize}
Using this algorithm, a subset of basis functions is selected. For one data point, the maximum number of basis functions $B_\mathbf{v}$ selected is equal to size of the set $\mathbf{\alpha}$. However, with multiple data points sampled the size of the final functionally-independent subset can be larger than size of the set $\mathbf{\alpha}$.

The difference in the Langbein algorithm's performance with random data points and training data was investigated in preliminary work. Random data points more effectively sample the space of possible values, and were subsequently used in the algorithm as default. This also means that the subset of points chosen by the Langbein algorithm is independent from the training set used.

\subsection{Cartesian formulation of tensors and enumeration of invariants as graphs}
\label{sec:cartesian_graphs}

\subsubsection{Cartesian Tensors}
In the companion paper, the authors generate a flexible basis by computing the Cartesian moment tensors~\eqref{momentTensor} as projections of the functions to the monomial basis and deriving their irreducible tensor decomposition by subtracting their traces~\cite{bujack2025flexible}. The \emph{irreducible tensors} $T^{\ell}_{\vec{\mathbf{\alpha}}} (\mathbf{r})$~\cite{Coope1965} are the traceless fully symmetric tensors. They are the Cartesian equivalent of the spherical harmonics. They form the building blocks of the flexible basis and are multiplied and contracted to form invariants. In this paper, we deal with spherical functions only. In contrast to general volumetric functions, the traces of the moment tensors are linear multiples of lower-rank moments tensors and don't need to be considered. The irreducible decomposition simplifies to a single full-rank irreducible tensor. Therefore, instead of removing the trace of moments in the monomial basis, we can directly compute the irreducible moment tensor by projecting the function onto the traceless part of the monomial tensors $r^{\otimes \ell}$.
The first three irreducible tensors $T^{\ell}_{\vec{\mathbf{\alpha}}} (\mathbf{r})$, where $\vec{\alpha} = (\alpha_1,\alpha_2,...\alpha_\ell)$ take the shapes
\begin{align}
\begin{split}
T^{0}(\mathbf{r}) & =1\\ 
T_{\alpha_0}^{1}(\mathbf{r}) & =r_{\alpha_0}\\
T_{\alpha_0 \alpha_1}^{2}(\mathbf{r}) & =r_{\alpha_0}r_{{\alpha_1}}-\frac{1}{3} \delta_{{\alpha_0} {\alpha_1}} r^{2} \\
T_{{\alpha_0}{\alpha_1}{\alpha_2}}^{3}(\mathbf{r}) & =r_{\alpha_0}r_{{\alpha_1}}r_{{\alpha_2}}-\frac{1}{5}(\delta_{{\alpha_0}{\alpha_1}}r_{{\alpha_2}}+\delta_{{\alpha_0}{\alpha_2}}r_{{\alpha_1}}+\delta_{{\alpha_1}{\alpha_2}}r_{{\alpha_0}})r^{2}. 
\end{split}\label{eq:irreducible_tensors}
\end{align}

For a given atomic environment function $f(\mathbf{r})$ in terms of a direction $\hat{\mathbf{r}}$, the Cartesian projection generates a tensor functional on the environment,
\begin{equation}
    \boldsymbol{f}^\ell_{\vec{\alpha}}  = \int f(\hat{\mathbf{r}}) T^\ell_{\vec{\alpha}} (\hat{\mathbf{r}}) d \hat{\mathbf{r}},
\end{equation}
where $\vec{\alpha} = (\alpha_0,\alpha_1, ...,\alpha_{\ell-1})$ represents the full set of Cartesian indices. Therefore $f^\ell_{\vec{\alpha}}$ is equivalent to the moment tensor ${}^\ell M$~\eqref{momentTensor} of a function over the sphere after subtraction of its trace. In the companion paper, the authors use the notation ${}^\ell_\ell H$ to refer to it. Each index $\alpha_i$ transforms under rotation as a vector, and the full tensor rotates with a product of rotation matrices, one for each Cartesian index. The notion of reducible and irreducible tensors describes whether or not the action of the full rotation matrix can be decomposed into functionally independent subspaces (reducible) or not (irreducible). From this, it can be determined that irreducible components are fully traceless and fully symmetric. The spherical tensors of angular momentum $\ell$ indexed by $m$, and irreducible Cartesian rank-$\ell$ tensors indexed by $\vec{\alpha}$, differ only by a change of basis and normalization; they can contain exactly the same information.

\subsubsection{Invariants and graphs}
One aspect of the Cartesian approach is that it does not require Clebsh-Gordon coefficients to construct invariants, instead, invariants can be constructed by ordinary contraction over the indices:  when all indices are contracted, the full expression is invariant. The trade-off of not having Clebsh-Gordon coefficients is that partial contractions with $\ell$ free indices are not always irreducible rank $\ell$ tensors because they may not be traceless and symmetric. For example, we can form invariants out of a Cartesian 2-tensor $\boldsymbol{f}^2 = f^2_{\alpha\beta}$ by following recipes:
\begin{align}
\begin{split}
I(f) &= \sum_{\alpha\beta} f^2_{\alpha\beta} f^2_{\alpha\beta}  = \mathrm{Tr}\left[ \boldsymbol{f}^2\cdot \boldsymbol{f}^2 \right]\\
J(f) &= \sum_{\alpha\beta\gamma} f^2_{\alpha\beta} f^2_{\beta\gamma} f^2_{\gamma\alpha} = \mathrm{Tr}\left[ \boldsymbol{f}^2 \cdot \boldsymbol{f}^2 \cdot \boldsymbol{f}^2 \right]\\
K(f) &= \sum_{\alpha\beta\gamma\delta} f^2_{\alpha\beta} f^2_{\beta\gamma} f^2_{\gamma\delta} f^2_{\delta\alpha} = \mathrm{Tr}\left[ \boldsymbol{f}^2 \cdot \boldsymbol{f}^2  \cdot \boldsymbol{f}^2 \cdot \boldsymbol{f}^2 \right]
\end{split} \label{eq: cartesian 2 tensors}
\end{align}

Here, suitable for the $\ell=2$ tensor,  $\mathrm{Tr}\left[\dots\right]$ denotes the matrix trace operation and $\cdot$ the matrix multiplication operation. The order of the contraction indices does not matter because the tensors are symmetric.  When considering tensors of varying ranks, we need not explicitly label $\ell$ for each tensor, because this is given by the number of indices. This allows us to associate invariant contractions with multi-graphs, i.e. graphs where more than one edge may be associated with each pair of nodes, as follows:
\def\invI{
\begin{tikzpicture}[baseline=-3pt,node distance={5mm}, thin, main/.style = {draw, circle}] 
\node[main] (1) {}; 
\node[main] (2) [right of=1]{}; 
\draw (1.20) -- (2.160); 
\draw (1.340) -- (2.200);
\end{tikzpicture}
}
\def\invJ{\begin{tikzpicture}[baseline=-7pt,node distance={5mm}, thin, main/.style = {draw, circle}] 
\node[main] (1) {}; 
\node[main] (2) [below left of=1] {}; 
\node[main] (3) [below right of=1] {};
\draw (1) -- (2); 
\draw (1) -- (3); 
\draw (3) -- (2); 
\end{tikzpicture} 
}
\def\invK{\begin{tikzpicture}[baseline=-10pt,node distance={5mm}, thin, main/.style = {draw, circle}] 
\node[main] (1) {}; 
\node[main] (2) [below of=1] {}; 
\node[main] (3) [right of=1] {};
\node[main] (4) [below of=3] {};
\draw (1) -- (2); 
\draw (1) -- (3); 
\draw (3) -- (4); 
\draw (4) -- (2); 
\end{tikzpicture} 
}
\begin{align}
I(f) &= \invI  \\
J(f) &= \invJ  \\
K(f) &= \invK . 
\end{align}
This kind of diagrammatic representation, referred to as tensor diagrams or Penrose diagrams~\cite{taylor2024introduction}, has a variety of uses and is used to visualize moment invariants and remove redundancies~\cite{suk2004graph}. The graph captures the notion that the order of indices within tensors is irrelevant; edges need not be labeled. Tracelessness means that graphs with self-loops must be equal to zero, and can therefore be ignored. Another advantage of the graph approach is to easily avoid examining invariants that factor into a product of lower-order invariants. In the graph formulation, this corresponds to a graph that has multiple connected components. This kind of redundancy has been identified previously by the authors of NICE in the spherical tensor formulation~\cite{Nigam2020nice}.  In the present work, we shall consider primarily the invariants of a single function $f(\hat{\mathbf{r}})$, and in this context there is no need to label nodes; nodes with different degree $\ell$ correspond to precisely to the tensor $\boldsymbol{f}^\ell$.

\subsubsection{Identifying a flexible basis of invariants}
Similarly to ACE, Cartesian graph invariants sets can scale combinatorically, and the problem of redundant invariants emerges even for simple cases such as the functions in Eq.~\ref{eq: cartesian 2 tensors}. 
First we consider regular graphs, i.e., graphs in which all nodes have the same number of edges. They correspond to \emph{pure} invariants~\cite{bujack2025flexible}, i.e., invariants that are contractions of powers of a single irreducible moment tensor.
It is well known that a traceless, symmetric 2-tensor contains only two invariant degrees of freedom, which can be seen from its eigendecomposition. Somewhat more algebra is required to identify the relation between the three second-rank pure graph invariants from~\eqref{eq: cartesian 2 tensors}, which is
\begin{align}
    K(f) &=   \frac{1}{2} I(f)^2 \\
    \invK &= \frac{1}{2} \left( \invI \right) ^2 .
\end{align}
However, a matrix decomposition does not apply straightforwardly to rank $\ell=3$ tensors, and becomes conceptually more difficult for higher-rank tensors. Nonetheless, they also contain relations. An example relation among the set of $\ell=3$ graphs is
\def\IL{\begin{tikzpicture}[baseline=-1mm,node distance={5mm}, thin, main/.style = {draw, circle}] 
\node[main] (1) {}; 
\node[main] (2) [right of=1] {}; 
\draw (1.40) -- (2.140); 
\draw (1.0) -- (2.180); 
\draw (1.320) -- (2.220);
\end{tikzpicture} }
\def\IM{\begin{tikzpicture}[baseline=-3.5mm,node distance={5mm}, thin, main/.style = {draw, circle}] 
\node[main] (1) {}; 
\node[main] (2) [right of=1] {}; 
\node[main] (3) [below of=1]{}; 
\node[main] (4) [below of=2] {}; 
\draw (1.20) -- (2.160); 
\draw (1.340) -- (2.200);
\draw (1.270) -- (3.90); 
\draw (2.270) -- (4.90); 
\draw (3.20) -- (4.160); 
\draw (3.340) -- (4.200); 
\end{tikzpicture} }
\def\IN{\begin{tikzpicture}[baseline=-3.5mm,node distance={5mm}, thin, main/.style = {draw, circle}] 
\node[main] (1) {}; 
\node[main] (2) [right of=1] {}; 
\node[main] (3) [below of=1]{}; 
\node[main] (4) [below of=2] {}; 
\draw (1.0) -- (2.180); 
\draw (1.270) -- (3.90); 
\draw (2.270) -- (4.90); 
\draw (3.0) -- (4.180); 
\draw (1) -- (4);
\draw (2) -- (3);
\end{tikzpicture}}
\begin{equation}
    \IM + \IN =  \frac{1}{2} \left( \IL \right) ^2 .
\end{equation}

The $\ell=3$ tensor corresponds to a particular function on the sphere. There are manifestly $2 \ell + 1 = 7$ components to this function, however, three degrees of freedom in these components determine the orientation of the function (parameterized, for example, by three Euler angles). 

For regular graphs, all that is needed is to use the Langbein approach to identify graphs that contain functionally-independent information until a set of $2\ell -2 $ invariants has been found. There are exceptions for the number of invariants for $\ell=0$ and $\ell = 1$ owing to the reduced degrees of freedom necessary to orient scalars and vectors. In total, the number of pure invariants that can be extracted from the tensor representations up to an order $\ell$ is quadratic in $\ell$. The complete theorems and a table summarizing the degrees of freedom of functionally independent invariants can be found in the companion paper. There it is also shown that no complete set can be generated using pure invariants only.

Finding information in the set of mixed invariants, i.e., contractions that contain tensors of differing rank, might appear to be a more daunting task, owing to the variety of non-regular graphs. However, with the pure invariants taken care of, the mixed invariants need only perform a simpler task: for each pair of orders $\ell$ and $\ell' \ne \ell$, the mixed invariants only need to determine the relative orientation of the functions represented by $\boldsymbol{f}^\ell$ and $\boldsymbol{f}^{\ell'}$, which requires at most three invariants. There are again exceptions to this count when one order $\ell$ is small, in particular, there are no independent mixed invariants involving $\ell=0$, which is in itself a graph, and only two degrees of freedom are needed to describe the orientation of a pattern relative to a vector ($\ell=1$).  

Bujack \etal proved that a basis of invariants can be generated by taking all pure invariants, then choosing one order $\ell_0$ whose moment tensor is guaranteed to be non-zero robustly across all admissible functions and adding three mixed invariants between it and each other tensor~\cite{bujack2022systematic}. While such a non-zero tensor can be chosen in some pattern recognition applications, this is not possible in our application. For a function whose $\ell_0$ tensor is zero, all mutual-orientation information is lost and the basis is no longer complete.
To reliably guard against the problem of a-priori-unknown vanishing tensors, one must include mixed invariants that connect all orders $\ell$ to all other orders $\ell'$. This so-called minimal flexible set~\cite{bujack2025flexible} is not a basis because it is not functionally independent. But it is the smallest possible set that is complete for any a-priori-unknown function. While this comes at the cost of a larger number of invariants, this scaling is quadratic and therefore asymptotically equivalent to any invariant basis for spherical functions.

\subsubsection{The Feature Sets in Practice}
Even though these results guarantee optimal feature sets in theory, in practice other criteria play a role. For example, descriptors vary in their evaluation speed. Further, descriptors with lower polynomial order---a function of the order of the moments and number of factors in the tensor product---are less prone to loss of precision in floating-point calculations.

One way to reduce the theoretically-optimal feature sets to feature sets that work efficiently and robustly in practice is by explicitly limiting the total invariant polynomial order, or the number of factors in the products, and using only the subset of the optimal set that falls below these limits. 
Alternatively, this limit can be applied before Langbein's algorithm is executed. The advantage of this is that fewer candidates need to be tested. The disadvantage is that the stopping criterion is a-priori unknown.

With these guidelines, we have performed a numerical search for a flexible set of Cartesian tensor invariants by enumerating graphs, constructing random tensors, and applying the Langbein test for independence. Table~\ref{tab:invariants_in_HIP-HOP-NN} lists a minimal flexible set up to a number of factors $n=4$ and moment order $\ell=3$, which in this special case happens to be equivalent to limiting the total descriptor polynomial order to $m=12$. The result includes one scalar graph, three tensor-norm graphs containing exactly two nodes each, two further pure invariants, and eight mixed invariants. This set can characterize all of the available information in a single function up to $\ell=3$ while using at most four terms in an invariant.

We use the minimal flexible set from Example~9 in the companion paper~\cite{bujack2025flexible} and restrict it to a total descriptor polynomial order 12. This reduces the original complete set of eight pure invariants to six. The seven mixed invariants are not affected. This leads to a total of 13 invariants-which coincides with the size of a basis for these tensors.

\begin{table}
\label{tab:invariants}
\begin{tabular}{r|c|c|c|c|c|c|}

\# & $\ell_\text{max}$ & $n$ & Terms                                     & Indices         & Graph \\
\hline
0      & 0     & 1    & $0                        $& -               &  
\begin{tikzpicture}[baseline=0pt,node distance={5mm}, thin, main/.style = {draw, circle}] 
\node[main] (1) {}; 
\end{tikzpicture}      \\
\hline
1      & 1     & 2   & $1^2                      $& 0,0             &     
\begin{tikzpicture}[baseline=0pt,node distance={5mm}, thin, main/.style = {draw, circle}] 
\node[main] (1) {}; 
\node[main] (2) [right of=1] {}; 
\draw (1.0) -- (2.180); 
\end{tikzpicture}   \\
\hline
2      & 2    & 2     & $2^2                      $& 01,01           &  
\begin{tikzpicture}[baseline=0pt,node distance={5mm}, thin, main/.style = {draw, circle}] 
\node[main] (1) {}; 
\node[main] (2) [right of=1] {}; 
\draw (1.20) -- (2.160); 
\draw (1.340) -- (2.200);
\end{tikzpicture}      \\
3      & 2     & 3     & $2^3                      $& 01,12,02        &    
\begin{tikzpicture}[baseline=-\fsize,node distance={5mm}, thin, main/.style = {draw, circle}] 
\node[main] (1) {}; 
\node[main] (2) [below left of=1] {}; 
\node[main] (3) [below right of=1] {};
\draw (1) -- (2); 
\draw (1) -- (3); 
\draw (3) -- (2); 
\end{tikzpicture} 
  \\
\hline
4      & 3     & 2     & $3^2                      $& 012,012         &   
\begin{tikzpicture}[baseline=-\fsize,node distance={5mm}, thin, main/.style = {draw, circle}] 
\node[main] (1) {}; 
\node[main] (2) [right of=1] {}; 
\draw (1.40) -- (2.140); 
\draw (1.0) -- (2.180); 
\draw (1.320) -- (2.220);
\end{tikzpicture}      \\
5      & 3     & 4     & $3^4                      $& 012,013,245,345 & 
\begin{tikzpicture}[baseline=-\fsize,node distance={5mm}, thin, main/.style = {draw, circle}] 
\node[main] (1) {}; 
\node[main] (2) [right of=1] {}; 
\node[main] (3) [below of=1]{}; 
\node[main] (4) [below of=2] {}; 
\draw (1.20) -- (2.160); 
\draw (1.340) -- (2.200);
\draw (1.270) -- (3.90); 
\draw (2.270) -- (4.90); 
\draw (3.20) -- (4.160); 
\draw (3.340) -- (4.200); 
\end{tikzpicture} 
      \\
\hline
6      & 2     & 3     & $1^2 2^1                    $& 0,1,01          &  
\begin{tikzpicture}[baseline=0pt,node distance={5mm}, thin, main/.style = {draw, circle}] 
\node[main] (1) {}; 
\node[main] (2) [right of=1] {}; 
\node[main] (3) [right of=2] {};
\draw (1.0) -- (2.180); 
\draw (2.0) -- (3.180); 
\end{tikzpicture}     \\
7      & 2     & 4     & $1^2 2^2                  $& 0,1,02,12      &    
\begin{tikzpicture}[baseline=0pt,node distance={5mm}, thin, main/.style = {draw, circle}] 
\node[main] (1) {}; 
\node[main] (2) [right of=1] {}; 
\node[main] (3) [right of=2] {};
\node[main] (4) [right of=3] {};
\draw (1.0) -- (2.180); 
\draw (2.0) -- (3.180); 
\draw (3.0) -- (4.180); 
\end{tikzpicture} 
   \\
   
\hline
8      & 3     & 4     & $1^3 3^1                    $& 0,1,2,012       &    
\begin{tikzpicture}[baseline=-\fsize,node distance={5mm}, thin, main/.style = {draw, circle}] 
\node[main] (1) {}; 
\node[main] (2) [above left of=1] {}; 
\node[main] (3) [above right of=1] {};
\node[main] (4) [below of=1] {};
\draw (1) -- (2); 
\draw (1) -- (3); 
\draw (1) -- (4); 
\end{tikzpicture} 
  \\
9      & 3     & 4     & $1^2 3^2                  $& 0,1,023,123    &  
\begin{tikzpicture}[baseline=0pt,node distance={5mm}, thin, main/.style = {draw, circle}] 
\node[main] (1) {}; 
\node[main] (2) [right of=1] {}; 
\node[main] (3) [right of=2] {};
\node[main] (4) [right of=3] {};
\draw (1.0) -- (2.180); 
\draw (2.20) -- (3.160); 
\draw (2.340) -- (3.200); 
\draw (3.0) -- (4.180); 
\end{tikzpicture}\\
\hline
10     & 3     & 3     & $2^1 3^2                    $& 01,023,123      &    
\begin{tikzpicture}[baseline=-\fsize,node distance={5mm}, thin, main/.style = {draw, circle}] 
\node[main] (1) {}; 
\node[main] (2) [below left of=1] {}; 
\node[main] (3) [below right of=1] {};
\draw (1) -- (2); 
\draw (1) -- (3); 
\draw (3.160) -- (2.20); 
\draw (3.200) -- (2.340); 
\end{tikzpicture}    \\
11     & 3     & 4     & $2^2 3^2                  $& 01,02,134,234  &    
\begin{tikzpicture}[baseline=-\fsize,node distance={5mm}, thin, main/.style = {draw, circle}] 
\node[main] (1) {}; 
\node[main] (2) [right of=1] {}; 
\node[main] (3) [below of=1]{}; 
\node[main] (4) [below of=2] {}; 
\draw (1.0) -- (2.180); 
\draw (1.270) -- (3.90); 
\draw (2.270) -- (4.90); 
\draw (3.20) -- (4.160); 
\draw (3.340) -- (4.200); 
\end{tikzpicture}    \\
12     & 3     & 4     & $2^2 3^2                  $& 01,23,014,234  &  
\begin{tikzpicture}[baseline=0pt,node distance={5mm}, thin, main/.style = {draw, circle}] 
\node[main] (1) {}; 
\node[main] (2) [right of=1] {}; 
\node[main] (3) [right of=2] {};
\node[main] (4) [right of=3] {};
\draw (1.20) -- (2.160); 
\draw (1.340) -- (2.200); 
\draw (2.0) -- (3.180); 
\draw (3.20) -- (4.160); 
\draw (3.340) -- (4.200); 
\end{tikzpicture} 
 
\end{tabular}
\centering
\begin{caption}{Flexible invariant basis of moment order $\ell=3$ reduced to $n\leq4$ terms. The body-order associated with with these graphs is given by $n+1$ and the polynomial degree is the sum of the degrees $\ell$ in each factor of the graph. Invariants 1-5 are pure and invariants 6-12 are mixed.}\label{tab:invariants_in_HIP-HOP-NN}
\end{caption}
\end{table}

\subsection{Incorporating flexible bases into atomistic neural networks \label{sec:hip-hop-methods}}

We start with the general form of a HIP-NN interaction layer. Atomic features $z_{i,a}$, where $i$ indexes atoms and $a$ indexes features, are transformed using the functional form

\begin{equation}
{z'}_{i,a}=f\left(\mathcal{I}_{i,a}(z,\mathbf{r})+\sum_{b}W_{ab}z_{i,b}+B_{a}\right).\label{eq:hipnn-layer}
\end{equation}

The weight matrix $W_{ab}$ describes a feature transformation that is local to each atom between features $a$ and $b$, with a bias vector $B$. The activation function $f$ acts point-wise on each element. The interaction function $\mathcal{I}$ specifies how features in the environment can affect a particular atom $i$. This form is rather general, and with specific functional forms for $\mathcal{I}$, one can arrive at many kinds of models from the literature. In HIP-NN, the interaction function is an explicit product of a radial representation $s^\nu(r_{ij})$ and the features $z_j$ on neighboring atoms with a learnable weight tensor (in the multi-dimensional array sense) $V^\nu_{ab}$:

\begin{equation}
\begin{aligned}
\mathcal{I}_{i,a}(z,r)^{\textrm{HIP-NN}} &= \sum_j m_{ij,a} \\
&=\sum_{j, b}v_{ab}(r_{ij})z_{j,b} \\
&= \sum_{j,b,\nu}  V_{ab}^\nu s^\nu(r_{ij}) z_{j,b}  .
\end{aligned} \label{eq:m_def}
\end{equation}
Here we define an intermediate message $m_{ij,a}$, sent from neighbor $j$ to atom $i$ on feature index $a$, as well as a variable weight function $v_{ab}(r_{ij})$, with $r_{ij} = |\boldsymbol{r}_i-\boldsymbol{r}_j|$. To expand on this framework to include directional information, it is natural to consider an environment function $\mathcal{E}_{i,a}(\hat{\mathbf{r}})$ which represents the messages received by an atom at any direction $\hat{r}$

\begin{equation}
\mathcal{E}_{i,a}(\hat{\mathbf{r}})=\sum_{j}\delta(\hat{\mathbf{r}}-\hat{\mathbf{r}}_{ij})m_{ij,a}.\label{eq:env_func}
\end{equation}

The message distribution can be broken down into irreducible tensor components using irreducible tensors $\boldsymbol{T}^{\ell}$ of Eq~\ref{eq:irreducible_tensors}. This generates environment tensors as projections of the environment function onto each Cartesian irreducible tensor:
\begin{equation}
    \boldsymbol{\mathcal{E}}_{i,a}^{\ell}=\int\mathbf{T}^{\ell}(\hat{\mathbf{r}})\mathcal{E}_{i,a}(\hat{\mathbf{r}})\,d^{2}\hat{\mathbf{r}}.
\end{equation}

In this notation, the original HIP-NN interaction is simply a single term in a series of environment tensors: 
\begin{equation}
\mathcal{I}_{i,a}^{\textrm{HIP-NN}}= \boldsymbol{\mathcal{E}}^{0}_{i,a}.
\end{equation}

Previously introduced was the notion of HIP-NN-TS, that is, HIP-NN with Tensor Sensitivities, which mixes environment tensors of different orders $\ell$ by taking their norms and mixing parameters $t^{\ell}_a$:
\begin{equation}
\mathcal{I}_{i,a}^{\textrm{HIP-NN-TS}}=\boldsymbol{\mathcal{E}}^{0}_{i,a}+t^{1}_a|\boldsymbol{\mathcal{E}}_{i,a}^{1}|+t^{2}_a|\boldsymbol{\mathcal{E}}_{i,a}^{2}|+\dots\, .
\label{eq:hip-vec-1}
\end{equation}

This same ansatz---that the tensor norms provide a useful but far cheaper subset of the full tensor information---was also used by SO3KRATES~\cite{Frank2024Euclidean,Kabylda2025molecular}. We now turn to the question of a strategy to integrate the invariants of section \ref{sec:cartesian_graphs} into a similar neural network architecture. Now, with a set of invariant functionals $\mathbb{I}^{k}[\cdot]$ which are indexed by $k$, operating as functions of the tensor components of the function as $\mathbb{I}^{k}(\boldsymbol{\mathcal{E}}^{0},{\boldsymbol{\mathcal{E}}^{1},\dots)}$, we can consider a set of scalar features $\mathbb{I}^k_{i,a}$ generated by these functionals:
\begin{equation}
    \mathbb{I}^k_{i,a} = \mathbb{I}^{k}[\mathcal{E}_{i,a}(\hat{\mathbf{r}})] .
\end{equation}
That is, the invariants are applied channel by channel, similarly to the tensor sketch as outlined by Darby et al.~\cite{darby2023tensor}  The invariants operating channel-wise do not heavily constrain the interaction because these channels are already  parameterized mixtures of the underlying features $z$, mixed using the tensor $V_{ab}^\nu$ of eq.~\ref{eq:m_def}. This is evident by performing the integration over the sphere and expanding the definition of $\boldsymbol{\mathcal{E}}_{i,a}^{\ell}$ in terms of scalar components:
\begin{equation}
    \mathcal{E}_{i,a,\vec{\alpha}}^{\ell} = \sum_{j,b,\nu}  V_{ab}^\nu T_{\vec{\alpha}}^{\ell}(\hat{\mathbf{r}}_{ij}) s^\nu(r_{ij}) z_{j,b}  ,
\end{equation}
where $\vec{\alpha} = (\alpha_0,\alpha_1,...,\alpha_{\ell-1}$) represents the $\ell$ Cartesian indices associated with a rank-$\ell$ tensor. We choose these functions $\mathbb{I}$ to be a flexible basis of invariant polynomials in accordance with the search from section~\ref{sec:cartesian_graphs}. 

While these features are scalar, and it is possible to use them in an arbitrary neural network layer, early experiments demonstrated a difficulty with handling especially high-order invariants. The difficulty is that the order of magnitude of high-order invariants can easily be larger or smaller than those of low order invariants; a fourth order polynomial can easily have a magnitude far higher or lower than a 2nd order polynomial. Naturally, the magnitude of the environmental features will vary in varying environments, and as such, changes in the magnitudes of the tensors will have a far greater effect on the magnitude of higher-order invariants than lower-order ones. A remedy to this  is \emph{group normalization}, which adjusts groups of activations relative to their own mean and standard deviation~\cite{wu2018group}. We therefore apply group normalization where each group corresponds to its own invariant graph, $k$, across the features, $a$, to compensate for the affect of different invariants being larger or smaller, yielding activations $\tilde{\mathbb{I}}^k_{i,a}$:
\begin{equation}
    \tilde{\mathbb{I}}^k_{i,a} = \mathrm{GroupNorm}\left<k\right>(\mathbb{I}^k_{i,a}) .
\end{equation}
We then mix these features to define a new interaction function. Because the flexible set of invariant polynomials allow the computations at large body-order, we refer to this version of HIP-NN using high-order polynomials as HIP-HOP-NN:
\begin{equation}
\mathcal{I}_{i,a}^{\textrm{HIP-HOP-NN}} = \sum_{b,n} \mathbb{W}_{ab}^{k} \tilde{\mathbb{I}}^{k}_{i,b} .
\end{equation}

\section{Results}
\label{sec:Results}

\subsection{ACE Basis Reduced with the Langbein Algorithm}

\begin{figure*}[htb]
    \centering
    \includegraphics[width=6.5in]{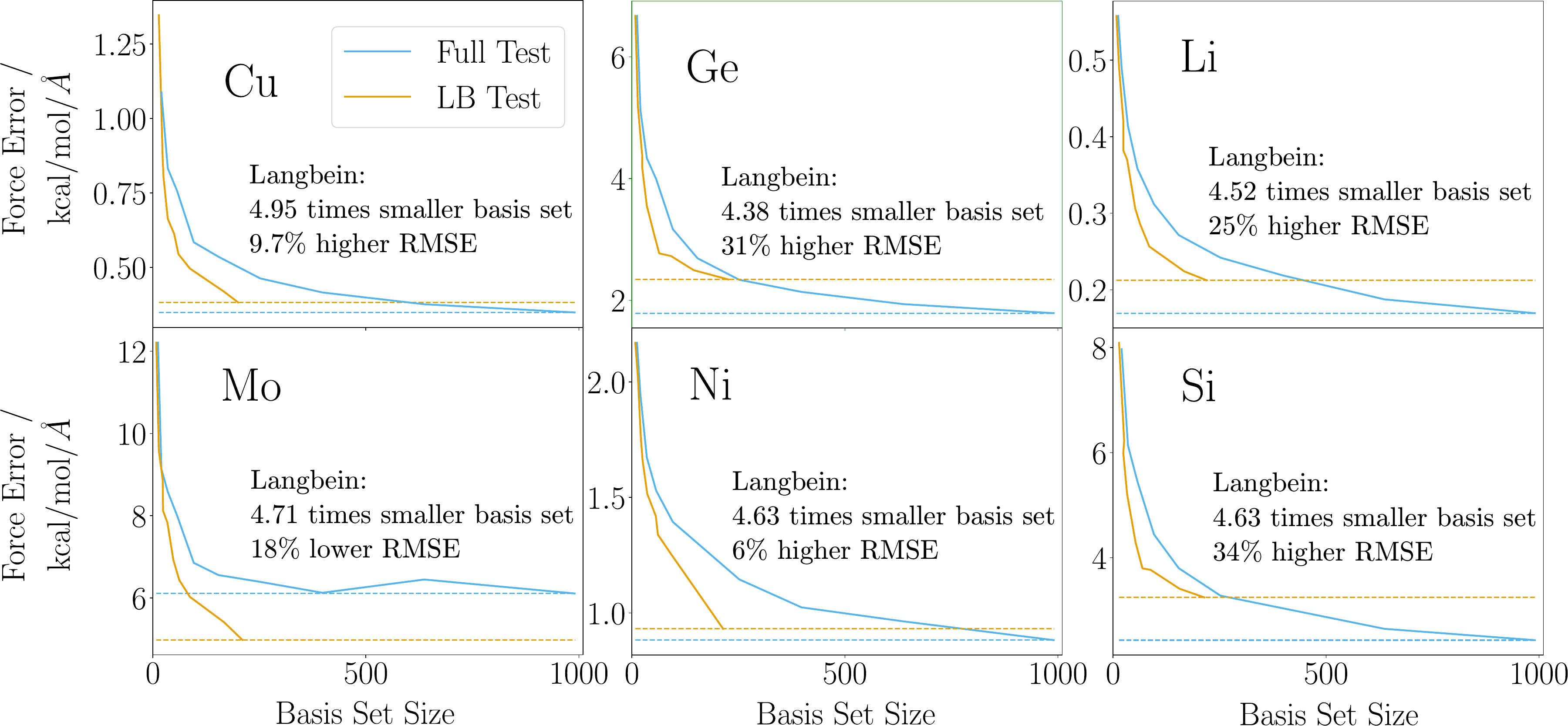}
    \caption{The change in test set force error with the size of the basis set for the six materials present in Ref.~\citenum{Zuo2020}. A comparison of the full basis set and the basis set reduced with the Langbein (LB) algorithm is shown. The dashed lines show the lowest error achievable with each basis.  For a given number of basis elements, the LB basis set provides consistently better accuracy.}
    \label{fig:ACE_Force_Langbein}

\end{figure*}

To show the applicability of the ideas introduced, we now turn to the ACE basis which is constructed from spherical harmonics. The dataset introduced in Ref.~\citenum{Zuo2020} contains six elements (Li, Mo, Cu, Ni, Si, and Ge) and samples a range of structures with ground state, strained, slab, and AIMD-sampled structures present. The quantum mechanical (QM) calculations are carried out with density functional theory with the Perdew–Burke–Ernzerhof (PBE) generalized gradient approximation (GGA) exchange-correlation function and were performed with VASP. The Langbein algorithm was implemented in the ACEpotentials.jl code described in Ref.~\citenum{Witt2023}.

Figure \ref{fig:ACE_Force_Langbein} shows the force error for a given basis-set size with the full ACE basis and the ACE basis subset found with the Langbein algorithm. Consistently across basis-set sizes the Langbein algorithm finds a subset of the full ACE basis that has lower error than the comparable full ACE basis of the same size. This subset is selected without any knowledge of the training/testing data present.  For the three elements Cu, Mo, and Ni, the Langbein subset has a comparable  error to the full ACE basis set at a fraction of the size. Therefore, the basis set can be made several times smaller in these cases with little or no reduction in accuracy. Reducing basis-set size is important for performance, and removing unnecessary basis functions could improve extrapolation. For the remaining cases (Ge, Li, Si), the Langbein subset does not reach the minimum error obtained with the full ACE set. This is expected, as the ACE basis function will contain functions that can be recreated from one another but are useful for modeling the functional form of the potential energy surface. The energy errors for these cases are shown in Fig.~\ref{fig:ACE_Energy_Langbein} and show that the performance of the energy prediction does not decay by choosing the Langbein subset. 

\begin{figure}[htb]
    \centering
    \includegraphics[width=3.5in]{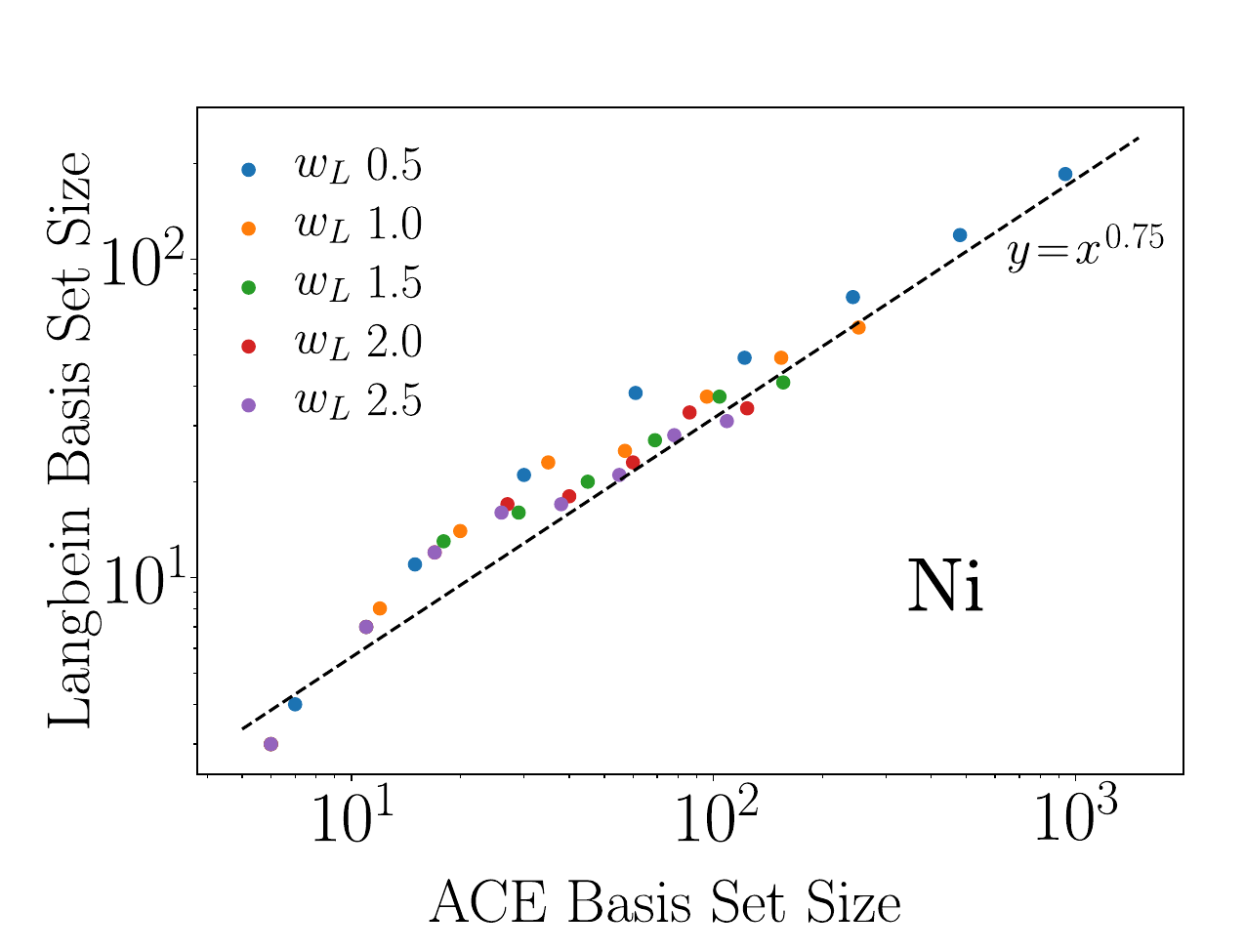}
    
    \caption{The reduction in the basis-set size with the Langbein algorithm for varying polynomial degrees and $w_L$ values. The degree of basis-set reduction is consistent and approximately follows a power law a with scaling exponent of 0.75. }
    \label{fig:ACE_Langbein_wl}
\end{figure}

Further properties of the Langbein subset were also investigated. In Fig.~\ref{fig:ACE_Langbein_wl}, the change in the length of the basis set is compared for Ni with the Langbein algorithm and the full ACE basis. The weighting of the radial component to the spherical component is also varied with the $w_L$ parameter. Across the different $w_L$ terms, the reduction in the size of ACE basis with the Langbein algorithm is relatively consistent. As the polynomial degree increases, and the size of the ACE basis increases, the relationship approximately follows a power law. 

We have seen how the Langbein subset of basis functions can produce excellent learning curves for the ACE basis. Importantly, this subset is chosen independently of the training data and is not influenced by it. Furthermore, the reduction of basis set size is consistent across different polynomial degrees and $w_L$ values. The use of techniques to construct optimal basis sets is now explored further with high order polynomial neural network.

\subsection{Neural network experiments}

\subsubsection{Methane Configurations Dataset}\label{subsec:methane dataset}

\begin{figure}[htb]
    \centering
    \includegraphics[width=3.5in]{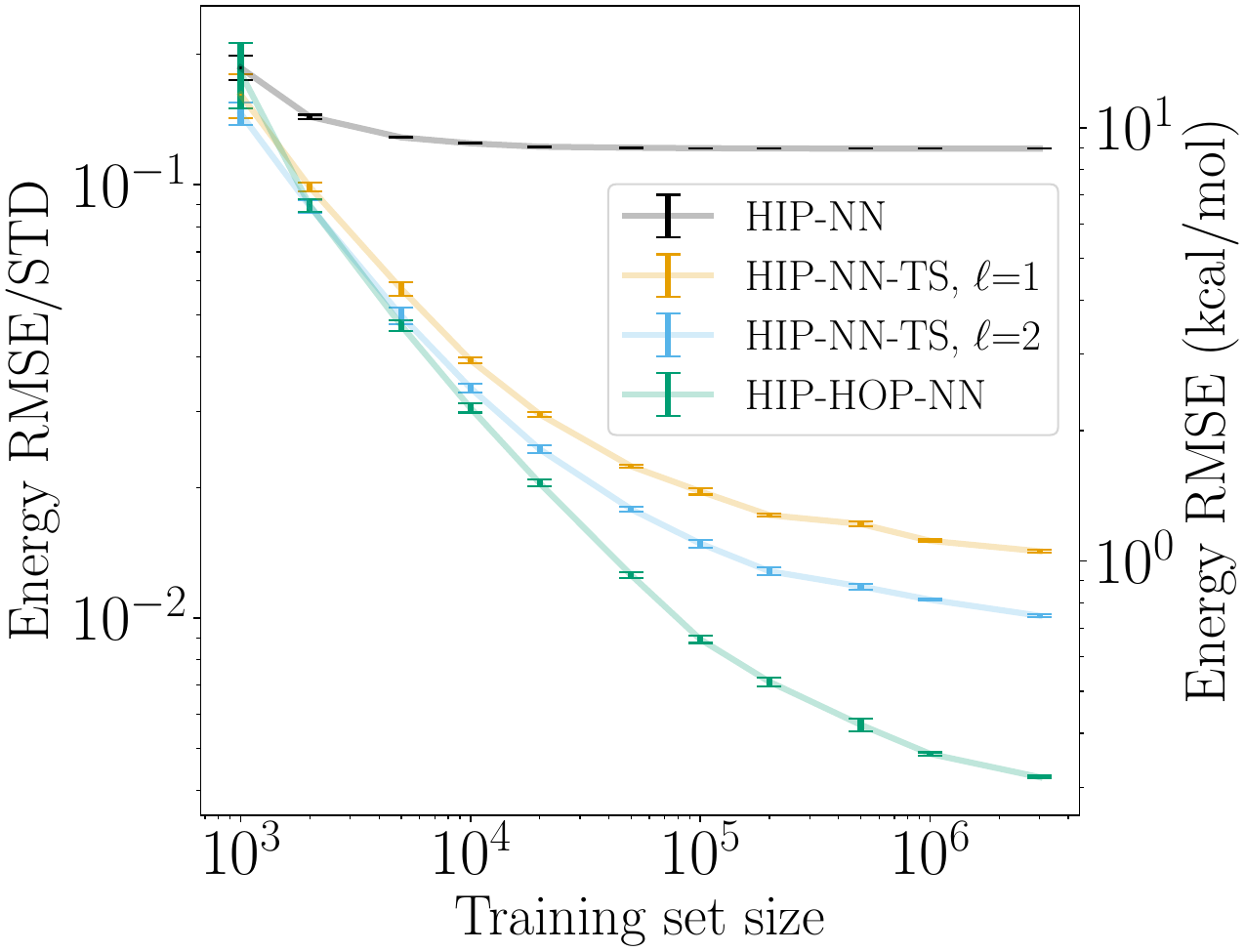}
    \caption{Model error versus training set size for different models trained on single-molecule methane configurations. For small dataset sizes, different HIP-NN architecture variants produce similar performance. As more and more data becomes available, HIP-HOP-NN is able to learn far more detail about geometries in the environment, significantly surpassing HIP-NN-TS and HIP-NN.}
    \label{fig:learning curves}

\end{figure}

The performance of HIP-HOP-NN against HIP-NN and HIP-NN-TS was explored using a dataset of randomly-displaced methane configurations~\cite{methane_dataset}. This dataset contains more than seven million configurations of methane with energies and forces calculated via density functional theory (DFT) using the PBE functional and cc-pVDZ basis set~\cite{methane_dataset}. A set of 80,000 configurations was reserved for testing---the  same configurations used for testing in Ref.~\citenum{Pozdnyakov2020}. From the rest of the configurations, 10 datasets were selected randomly for each of 11 sizes ranging from 1,000 to 3,000,000. Each of the 110 resulting datasets were used to train a HIP-NN model, two HIP-NN-TS models (with $\ell=1$ and $\ell=2$), and a HIP-HOP-NN model (with $n=4$ and $\ell=3$). The resulting energy RMSE values on the withheld training set were then calculated for each model. The results are displayed in Figure~\ref{fig:learning curves}. The plotted values represent the mean RMSE value among the 10 models trained for each network type and training set size and the error bars represent the standard error of the mean. All models used an identical, one interaction layer architecture, described in SI.2. 

The HIP-NN model performance stagnated quickly at around 20,000 training points with an energy root-mean-squared error (RMSE) of about 12\% of the data's standard deviation. HIP-NN models with two interaction layers were found to improve performance for larger dataset size, however we focus on just one interaction model comparisons for this system. Conversely, the HIP-NN-TS and HIP-HOP-NN models demonstrated improvement with each data size increase, with the two HIP-NN-TS variants achieving an RMSE of about 1.4\% and 1\% of the data standard deviation and the HIP-HOP-NN model achieving about 0.45\% on the largest training sets. The performance of these three network types are very similar when trained on the datasets of size at most 10,000. The advantage of the HIP-HOP-NN architecture can be seen best on the datasets of size 200,000 and greater, where the errors of the HIP-HOP-NN models are consistently half that of the HIP-NN-TS models. 

These results demonstrate the value of including higher-order features into atomic environment representations, especially when large training data sets are available. A related finding is shown in Ref.~\citenum{Pozdnyakov2020} Figure 4c. 

\subsubsection{QM7 Dataset}

To further understand the performance of HIP-HOP-NN on small datasets, we tested its performance against HIP-NN and HIP-NN-TS on the QM7 dataset~\cite{Montavon2013MachineSpace}, which contains 7,211 single-molecule configurations. The dataset includes atomization energies calculated using DFT for all small organic molecules that have up to seven total of C, N, O, S, and Cl, plus hydrogens. Small datasets remain a valuable benchmark test because in practical applications data availability may be limited due to computational cost or other constraints. As in the preceding section, we include two HIP-NN-TS models (with $\ell=1$ and $\ell=2$), one HIP-HOP-NN model (with $n=4$ and $\ell=3$), and one HIP-NN model in the comparison. A hyperparameter search was conducted using Bayesian optimization, detailed in section SI.3. 

The results of the hyperparameter search, displayed in Table~\ref{tab:qm7_results}, show that even with limited training data, the HIP-HOP-NN models are able to consistently outperform HIP-NN and HIP-NN-TS models, with a reduction in mean absolute error (MAE) of about 20\% compared to the next best architecture. An analysis of the effect of the the learning rate hyperparameter over the search showed that HIP-HOP-NN performance was more robust across a range of learning rates compared to HIP-NN (see section SI.3). We also find, in contrast to previous work~\cite{Chigaev2023}, that under thorough hyperparameter search, HIP-NN can out-perform HIP-NN-TS on this dataset.

\begin{table}
\begin{tabular}{|c|c|c|c|c|}
\hline
\textbf{Model}      & $\mathbf{n}$ & \boldmath$\ell$\unboldmath
& \textbf{MAE (kcal/mol)} & \textbf{RMSE (kcal/mol)} \\
\hline
HIP-NN     & 1   & 0      & 0.98 $\pm$ 0.11       & 2.37 $\pm$ 1.02  \\
\hline
HIP-NN-TS  & 2   & 1      & 1.02 $\pm$ 0.08       & 2.35 $\pm$ 0.75  \\
\hline
HIP-NN-TS  & 2   & 2      & 1.13 $\pm$ 0.06       & 2.21 $\pm$ 0.70  \\
\hline
HIP-HOP-NN & 4   & 3      & 0.79 $\pm$ 0.06       & 2.18 $\pm$ 0.89  \\
\hline

\end{tabular}
\centering
\begin{caption}{Mean and standard deviation of model error metrics, computed using ten random splits of the QM7 dataset. The same ten splits were used for each model. 80\% in each split was used for model training and the remaining 20\% to calculate these error metrics.}\label{tab:qm7_results}
\end{caption}
\end{table}

\subsubsection{QM9 Dataset}

The QM9 dataset contains equilibrium structures and DFT energies for 134k small organic molecules~\cite{Ramakrishnan2014QuantumMolecules}. The QM9 benchmark is one of the most widely used datasets for assessing ML models for molecular property prediction, however it cannot be directly used for interatomic potentials~\cite{Ramakrishnan2014QuantumMolecules}. We focus on modeling the atomization energy, calculated at the B3LYP/6-31G(2df,p) level of theory. The hyperparameters used are described in  SI.5. To further increase the accuracy of the model, a learning schedule for forces was introduced as described in SI.6. 

Table \ref{tab:qm9_comparison} shows the QM9 results for HIP-HOP-NN compared to other ML models. We can see that an ensembled HIP-HOP-NN model offers an accuracy comparable to the best ML models available. The increase in accuracy compared to prior results from HIP-NN-TS is due both to a change in the tensor set used and to the force training scheme and batch hierarchicality introduced. On this first benchmark test of HIP-HOP-NN compared to other models, HIP-HOP-NN shows excellent accuracy, demonstrating the advantages of the tensor construction introduced. 

\begin{table}
    \centering
\begin{tabular}{lccc}
Method & MAE \\
\toprule
Allegro~\cite{allegro} & 0.11 \\
HIP-HOP-NN Ensemble & 0.11 \\
HIP-NN-TS Ensemble & 0.12 \\
PaiNN~\cite{schutt2021equivariant} & 0.13 \\
ET~\cite{tholke2022torchmd} & 0.14 \\
HIP-HOP-NN & 0.14 \\
DimeNet$^{++}$~\cite{klicpera2020fast} & 0.15  \\
HIP-NN-TS  & 0.15 \\
HIP-NN Ensemble & 0.15 \\
DL-MPNN~\cite{DLMPNN}  & 0.17 \\
HIP-NN & 0.18 \\
PhysNet~\cite{Unke2019PhysNet:Charges} &  0.19 \\
\end{tabular}
\caption{The performance of HIP-HOP-NN compared to other published neural networks, as measured by MAE performance on atomization energy predictions for the QM9 dataset in units kcal/mol. 
Models without a citation given were trained in this work.}
\label{tab:qm9_comparison}
\end{table}

\subsubsection{ANI1x and COMP6 Testing}

The ANI-1x dataset contains the DFT energies and forces for 4.9 million conformations of organic molecules collected with active learning~\cite{Smith2018LessLearning, Smith2020TheMolecules}. Training to this dataset allows the assessment of HIP-HOP-NN as a transferable molecular potential capable of describing a wide range of chemical space. A subset of the ANI-1x dataset was chosen which also contained coupled cluster calculations and was approximately 10$\%$ of the total size. This subset was also examined using MACE~\cite{Kovacs2023}. The accuracy of the model can then be assessed on the COMP6 benchmark which contains molecules outside the original training dataset and serves as a test of extrapolation. 

The performance of HIP-HOP-NN is shown in Table.~\ref{tab:COMP6}. HIP-HOP-NN significantly outperforms the ANI-1x model, GM-NN, and NewtonNet across the COMP6 benchmark. The ANI-MD subset remains high-error for HIP-HOP-NN for energies, however this is primarily due to the large structures in this subset, and the force error remains excellent. For the two MACE models shown, HIP-HOP-NN is comparable in accuracy to the medium MACE model. This is an excellent level of accuracy, and again demonstrates the state-of-the-art performance achieved with HIP-HOP-NN. While the accuracy is lower than the large MACE model, it is worth noting that this model will be significantly slower and more memory intensive than HIP-HOP-NN. Similarly, while the accuracy of TensorNet is slightly higher than for HIP-HOP-NN, the TensorNet model was trained to the entirety of the ANI-1x dataset.

\begin{table}
\begin{tabular}{|rc|c|c|c|c|c|c|c|}
\hline
 & \textbf{} & \textbf{\begin{tabular}[c]{@{}c@{}}ANI-1x\\  \cite {Smith2018LessLearning}\end{tabular}} & \textbf{\begin{tabular}[c]{@{}c@{}}GM-NN\\  \cite {Zaverkin2023}\end{tabular}} & \textbf{\begin{tabular}[c]{@{}c@{}}MACE\\ 96-1 \cite {Kovacs2023}\end{tabular}} & \textbf{\begin{tabular}[c]{@{}c@{}}MACE\\ 192-2 \cite {Kovacs2023}\end{tabular}} & \textbf{\begin{tabular}[c]{@{}c@{}}NewtonNet\\  \cite {newtonnet}\end{tabular}} & \textbf{\begin{tabular}[c]{@{}c@{}}TensorNet\\  2L \cite {simeon2023}\end{tabular}} & \textbf{HIP-HOP} \\ \hline
\multicolumn{1}{|r|}{\multirow{2}{*}{\textbf{ANI-MD}}} & E & 3.40 & 3.83 & 2.81 & 3.25 & - & 1.61 & 3.55  \\ \cline{2-9} 
\multicolumn{1}{|r|}{} & F & 2.68 & 1.43 & 0.89 & 0.62 & - & 0.82 & 0.84 \\ \hline
\multicolumn{1}{|r|}{\multirow{2}{*}{\textbf{DrugBank}}} & E & 2.65 & 2.78 & 1.04 & 0.73 & - & 0.98 & 1.06 \\ \cline{2-9} 
\multicolumn{1}{|r|}{} & F & 2.86 & 1.69 & 0.70 & 0.47 & - & 0.75 & 0.75 \\ \hline
\multicolumn{1}{|r|}{\multirow{2}{*}{\textbf{GDB 7-9 }}} & E & 1.04 & 1.22 & 0.40 & 0.21 & - & 0.32 & 0.29 \\ \cline{2-9} 
\multicolumn{1}{|r|}{} & F & 2.43 & 1.41 & 0.54 & 0.34 & - & 0.53 & 0.52 \\ \hline
\multicolumn{1}{|r|}{\multirow{2}{*}{\textbf{GDB 10–13}}} & E & 2.30 & 2.29 & 0.88 & 0.53 & - & 0.83 & 0.79 \\ \cline{2-9} 
\multicolumn{1}{|r|}{} & F & 3.67 & 2.25 & 0.92 & 0.62 & - & 0.97 & 0.98 \\ \hline
\multicolumn{1}{|r|}{\multirow{2}{*}{\textbf{S66x8}}} & E & 2.06 & 2.95 & 0.69 & 0.39 & - & 0.62 & 0.68 \\ \cline{2-9} 
\multicolumn{1}{|r|}{} & F & 1.60 & 0.93 & 0.33 & 0.22 & - & 0.33 & 0.60  \\ \hline
\multicolumn{1}{|r|}{\multirow{2}{*}{\textbf{Tripeptides}}} & E & 2.92 & 3.06 & 1.18 & 0.79 & - & 0.92 & 1.04 \\ \cline{2-9} 
\multicolumn{1}{|r|}{} & F & 2.49 & 1.48 & 0.66 & 0.44 & - & 0.62 & 0.67 \\ \hline
\multicolumn{1}{|r|}{\multirow{2}{*}{\textbf{COMP6}}} & E & 1.93 & 2.03 & 0.76 & 0.48 & 1.45 & - & 0.64 \\ \cline{2-9} 
\multicolumn{1}{|r|}{} & F & 3.09 & 1.85 & 0.77 & 0.52 & 1.79 & - & 0.80 \\ \hline
\end{tabular}
\caption{A comparison of the COMP6 performance of HIP-HOP-NN with other state of the art MLIPs. The errors are given in kcal/mol for energies ($E$) and forces ($F$) are in kcal/mol/Å. The subsets of ANI-1x used for training are not consistent across the MLIPs shown. HIP-HOP-NN is trained on $\approx$10\% of the full ANI-1x dataset.}
\label{tab:COMP6}
\end{table}

\subsubsection{Computational Cost}

To test the computational efficiency of HIP-HOP-NN, we performed the speed tests for inference of small molecules discussed in Ref.~\citenum{simeon2023}. The results are shown in  Table~\ref{tab:speed} with comparisons to TensorNet and MACE-OFF~\cite{kovacs2023maceoff,simeon2023}. The TensorNet results used a NVIDIA GeForce RTX 4090 GPU whereas the MACE and HIP-HOP-NN used an NVIDIA A100 Tensor Core GPU with 80 GB HBM2 memory. For the large MACE-OFF model, the Factor IX calculation could not be performed due to memory limitations. The NVIDIA cuEquivariance implementation was not used for the speed tests shown. 

HIP-HOP-NN shows excellent performance across the system sizes. HIP-HOP-NN is faster than all of the MACE models and TensorNet-2L for the alanine dipeptide and chignolin. For the largest systems, DHFR and Factor IX, HIP-HOP is slower than the smallest MACE-OFF model but is significantly faster than the MACE-OFF medium model. Therefore, HIP-HOP-NN produces a model that is competitive with state-of-the-art models for both speed and computational performance. This highlights the benefits of using a specialized construction for a compact set of tensor expressions. 

\begin{table}
\begin{tabular}{|r|c|c|c|c|c|c|}
\hline
\textbf{Molecule} & \textbf{N} & \textbf{\begin{tabular}[c]{@{}c@{}}TensorNet \\ 2L\end{tabular}} & \textbf{\begin{tabular}[c]{@{}c@{}}MACE-OFF \\ small\end{tabular}} & \textbf{\begin{tabular}[c]{@{}c@{}}MACE-OFF \\ medium\end{tabular}} & \textbf{\begin{tabular}[c]{@{}c@{}}MACE-OFF \\ large\end{tabular}} & \textbf{\begin{tabular}[c]{@{}c@{}}HIP-HOP-NN\\ 256\end{tabular}} \\ \hline
\textbf{\begin{tabular}[c]{@{}r@{}}Alanine \\ Dipeptide\end{tabular}} & \textit{22} & 26.5 & 15.5 & 19.2 & 23.0 & 14.1 \\ \hline
\textbf{Chignolin} & \textit{166} & 26.9 & 15.5 & 24.4 & 68.7 & 13.7 \\ \hline
\textbf{DHFR} & \textit{2489} & 106.7 & 68.2 & 272.2 & 1025.6 & 95.6 \\ \hline
\textbf{Factor IX} & \textit{5807} & 248.6 & 143.1 & 610.2 & - & 210.7 \\ \hline
\end{tabular}
\caption{The inference time (in ms) for single molecules with batch size of one for HIP-HOP-NN compared to other state-of-the-art MLIPs.  HIP-HOP-NN shows performance that is highly competitive with other state-of-the-art MLIPs; HIP-HOP-NN can be up to three times faster than MACE-OFF medium while having comparable COMP6 accuracy. }
\label{tab:speed}
\end{table}

\section{Discussion and Conclusions}
\label{sec:conclusions}

In recent years, atomistic machine learning schemes have flourished based on the construction of ACE~\cite{AtomicClusterExpansion}.  ACE can describe all local-environment functions as linear combinations of the descriptors, but grows combinatorically under truncation. However, finding the optimal set of descriptors remains an ongoing challenge~\cite{Dusson2022}. We have described how flexible basis sets can be selected from this type of construction to form a complete basis which does not grow combinatorically under truncation. Flexibility can be probabilistically tested numerically using random inputs and examining the rank of the Jacobian of the features, an approach due to Langbein.

We applied the notion of flexibility to ACE as a feature selection algorithm, yielding a reduced set of features that empirically shows improved learning curves. The benefits of using this reduced ACE basis set in non-linear ACE approaches such as MACE or GRACE remains an open and interesting question~\cite{batatia2022mace, Bochkarev2024}. 

We also apply this method to Cartesian invariants of irreducible tensors, which can be expressed as graphs. This yields a compact set of tensor expressions that can be used within a neural network architecture, which we call HIP-HOP-NN. We implemented the architecture, based on a general flexible set of moment tensor invariants, up to a cutoff a number of factors of $n=4$ and moment order of $\ell=3$ in this work, thus extending the individual neurons to body-order 5. The performance of HIP-HOP-NN was tested against a variety of benchmarks. For the methane dataset, which probes the completeness of the model, HIP-HOP-NN showed significant improvement over its predecessors (HIP-NN and HIP-NN-TS) at large dataset sizes. The QM-7 dataset probes chemical complexity at small dataset sizes, containing just 7211 different molecules. This demonstrated that even for small dataset sizes, the approach is useful, implying that useful high-order many-body information can be found even in this limit. We further compared HIP-HOP-NN to results from the literature. Excellent accuracy was achieved for the QM9 test set, as well as the ANI-1x/COMP6 benchmarks, with performance comparable to the medium-size MACE models. We examined the computational cost of HIP-HOP-NN on biological systems of a variety of sizes, and found it to be slightly better than TensorNet, and approximately three times faster than MACE-OFF, when applied to systems of approximately a few thousand atoms. The code for HIP-HOP-NN has been incorporated into hippynn, an open-source library for atomistic machine learning, available at \url{https://www.github.com/lanl/hippynn}. Also included are example files which allow for training HIP-HOP-NN to ANI-1x and performing hyperparameter search over models for QM7.

Constructing efficient representations of atomic environments (or more generally, point clouds) is an ongoing challenge. We have shown how the notion of flexible basis sets can address this challenge by pruning features from larger sets. There are multiple routes for extension of this work. For one, other MLIP formulations might benefit from exploring the notion of flexible feature sets. Considering flexibility for the computation of covariant (also called equivariant) feature sets would further the range of possible applications. Additionally, the construction of pre-tabulated sets of flexible invariants for Cartesian and spherical tensors to very-high order would provide a valuable resource for the research community interested in this area; this is non-trivial as the number of invariants grows combinatorically with the number of factors in the invariant. Furthermore, exploring the benefits of including higher-order invariants into our models remains an open question.  While we have concentrated on atomistic systems, the overall approach is widely applicable well beyond the chemical sciences.

\section*{Acknowledgements}
The authors are thankful for productive conversations with Kipton Barros, Cristina Garcia Cardona, Ying Wai Li, Yen Ting Lin, Sakib Matin, and Pieter Swart. We would like to thank Kei Davis for his feedback on this manuscript.

We gratefully acknowledge the support of the U.S. Department of Energy through the LANL Laboratory Directed Research Development Program under project numbers 20250145ER, 20230293ER for this work. This research used resources provided by the LANL Institutional Computing (IC) Program and the CCS-7 Darwin cluster at LANL. LANL is operated by Triad National Security, LLC, for the National Nuclear Security Administration of the U.S. Department of Energy (Contract No. 89233218NCA000001).

\appendix
\clearpage

\section{Supplemental Information}
\label{sec:appendix}

\subsection{Functions on a sphere and spherical tensor features}
\label{sec:functions_on_sphere}

Here we consider the characterization of atomic environments. To simplify indices, we consider a single atom positioned at the origin in its local environment with atoms indexed by $j$, and ignore chemical degrees of freedom. The atom views its environment in terms of a function $f(\hat{\boldsymbol{r}})$ on the sphere of the unit vector $\hat{\boldsymbol{r}}$:
\begin{equation}
f(\hat{\boldsymbol{r}})=\sum_{j}\delta(\hat{\boldsymbol{r}}-\hat{\boldsymbol{r}}_{j}) \rho(r_j) \label{eq:sphere_func} ,
\end{equation}
where $\rho(r)$ is a radial weighting function describing the amplitude of the effect of a neighboring atom $j$, whose position $\vec{\boldsymbol{r}}_j = r_j \hat{\boldsymbol{r}}_j$ has direction $\hat{\boldsymbol{r}}_j$ and magnitude $r_j$. Many such functions can be produced by varying the density projection function $\rho$, which can take into account information such as the chemical species.

Abstract functions such as this are not simple to represent in a finite manner and to manipulate with a computer. The well-known spherical harmonics $Y^\ell_m(\hat{\boldsymbol{r}})=Y^\ell_m(\theta,\phi)$ provide a solution to this difficulty. They can be used to project a function into a \emph{spherical tensor} basis indexed by $(\ell,m)$, where $\ell \ge 0,\ -\ell \le m \le \ell$:
\begin{equation}
f^\ell_m  = \int f(\hat{\boldsymbol{r}}) Y^{\ell *}_{m}(\hat{\boldsymbol{r}}) d^2 \hat{\boldsymbol{r}}  .\label{eq:spherical_projection}
\end{equation}
The $Y^{\ell *}_m$ denotes the complex conjugate of $Y^{\ell}_m$. We note that the exact functional form of the spherical harmonics depends on choices of phase and normalization that are not relevant to the discussion here. Because the spherical harmonics form a complete basis, the function $f(\hat{\boldsymbol{r}})$ can be reconstructed from $f^\ell_m$:
\begin{equation}
f(\hat{\boldsymbol{r}})  = \sum_{\ell,m} f^{\ell}_{m} Y^\ell_m(\hat{\boldsymbol{r}}) \label{eq:spherical_reconstruction} .
\end{equation}
The great utility of using the spherical harmonics as a basis set derives from the fact that they decompose all functions according to \emph{irreducible subspaces} given by each value of $\ell$. This means that entire set of $f^\ell_m$ can be thought of as the direct sum of a set of tensors $\boldsymbol{f}^\ell$, each with $2\ell + 1$ components, that are each individually well-behaved and do not mix under rotation. This is of immense practical utility because it means that methods can be designed which cut off a function using a tensor expansion at finite $\ell$, while still preserving the exactness of all calculations with respect to three-dimensional rotations.

Multiple functions on the sphere, say, $f(\hat{\boldsymbol{r}})$ and $g(\hat{\boldsymbol{r}})$, can be used to yield new functions defined by the ordinary point-wise product $h(\hat{\boldsymbol{r}})=f(\hat{\boldsymbol{r}})g(\hat{\boldsymbol{r}})$~\cite{luo2024nabling}. This operation can also be applied to the tensor representation of the functions, an operation known as the Gaunt product~\cite{luo2024nabling}. The Gaunt product for the tensor representation $h^{l}_{m}$ for the function $h(\hat{\boldsymbol{r}})$ has the structure
\begin{equation}    
h^{l_3}_{m_3} = \sum_{\ell_1,m_1,{\ell_2},{m_2}} M^{\ell_1 \ell_2 \ell_3} C^{\ell_1 \ell_2;\ell_3}_{m_1 m_2;m_3} f^{\ell_1}_{m_1}  g^{\ell_2}_{m_2} ,\label{eq:gaunt_add}
\end{equation}
where $ C^{\ell_1 \ell_2;\ell_3}_{m_1 m_2;m_3}$ are the \emph{Clebsh-Gordon coefficients} (CG coefficients), and $f$ and $g$ are projections of different functions on the sphere (perhaps with differing radial weighting functions).
The factors $M^{\ell_1 \ell_2 \ell_3}$ are scalar in nature and can be viewed as normalization factors that serve to define the Gaunt product. Other generalized tensor-based product operations can be defined by using different normalization factors. 
The meaning of the CG coefficients are often a source of confusion, in part because of the many conventions of normalization and sign which can be found when studying the spherical harmonics. The role of the CG coefficients in the product of tensors can be considerably demystified by considering the computation of the $h^{\ell_3}_{m_3}$ in terms of $f(\hat{\boldsymbol{r}})$ and $g(\hat{\boldsymbol{r}})$, and inserting Eq.~\ref{eq:spherical_reconstruction} for both $f$ and $g$:
\begin{align}
    h^{\ell_3}_{m_3} &= \int h(\hat{\boldsymbol{r}}) Y^{\ell_3 *}_{m_3}(\hat{\boldsymbol{r}})  d^2 \hat{\boldsymbol{r}}\\
    &= \int f(\hat{\boldsymbol{r}}) g(\hat{\boldsymbol{r}}) Y^{\ell_3 *}_{m_3}(\hat{\boldsymbol{r}}) d^2 \hat{\boldsymbol{r}} \\
    &= \sum_{{\ell_1},{m_1}}   f^{{\ell_1}}_{m_1} \int  g(\hat{\boldsymbol{r}}) Y^{\ell_1}_{m_1}(\hat{\boldsymbol{r}}) Y^{\ell_3 *}_{m_3}(\hat{\boldsymbol{r}}) d^2 \hat{\boldsymbol{r}} \\
    &= \sum_{\ell_1,m_1,{\ell_2},{m_2}}   f^{\ell_1}_{m_1} g^{{\ell_2}}_{{m_2}} \int Y^{{\ell_1}}_{{m_1}}(\hat{\boldsymbol{r}})  Y^{{\ell_2}}_{{m_2}}(\hat{\boldsymbol{r}}) Y^{\ell_3 *}_{m_3}(\hat{\boldsymbol{r}}) d^2 \hat{\boldsymbol{r}} \label{eq:cg_integral_version} .
\end{align}
By recognizing Eq.~\ref{eq:cg_integral_version} as having precisely the same tensor structure as Eq.~\ref{eq:gaunt_add}, the CG coefficients can be viewed as the $(\ell_3,m_3)$-\emph{th} spherical expansion coefficient between the product of two harmonics $ Y^{{\ell_1}}_{{m_1}}(\hat{\boldsymbol{r}}) Y^{{\ell_2}}_{m_2}(\hat{\boldsymbol{r}})$:
\begin{equation}
    C^{\ell_1 \ell_2;\ell_3}_{m_1 m_2;m_3} = \frac{1}{M^{\ell_1 \ell_2 \ell_3}} \left( Y^{{\ell_1}}_{{m_1}} \times  Y^{{\ell_2}}_{{m_2}} \right)^{\ell_3}_{m_3} = \frac{1}{M^{\ell_1 \ell_2 \ell_3}} \int Y^{{\ell_1}}_{{m_1}}(\hat{\boldsymbol{r}}) Y^{{\ell_2}}_{{m_2}}(\hat{\boldsymbol{r}})  Y^{\ell_3 *}_{m_3}(\hat{\boldsymbol{r}}) d^2 \hat{\boldsymbol{r}} .
\end{equation}
Therefore, using $\{\ell\}$ to denote the space spanned by the $2 \ell+1 $ spherical harmonics with fixed $\ell$ and variable $m$, the CG coefficients effect a change of basis between the product space $\{\ell_1\} \boldsymbol{\otimes} \{\ell_2\}$ and the sum space $\bigoplus_{\ell_3=|\ell_1-\ell_2 |}^{| \ell_1+\ell_2 |} \{\ell_3\}$. The restriction of the sum space is because when $\ell_3 <|\ell_1-\ell_2 |$ or $\ell_3 >|\ell_1+\ell_2 |$, we have $C^{\ell_1 \ell_2;\ell_3}_{m_1 m_2;m_3}=0$.

A related product, more commonly used in the atomistic machine learning literature~\cite{thomas2018tensorfieldnetworksrotation,allegro} is the \emph{tensor product},
\begin{equation}    
h'^{l_3}_{m_3} = \sum_{\ell_1,m_1,{\ell_2},{m_2}} C^{\ell_1 \ell_2;\ell_3}_{m_1 m_2;m_3} f^{\ell_1}_{m_1}  g^{\ell_2}_{m_2} \label{eq:cg_add_2}
\end{equation}
In the tensor product, while $h'^{l_3}_{m_3}$ is a tensor expansion of a function on the sphere $h'(\hat{\boldsymbol{r}})$, there is not a simple expression for $h'(\hat{\boldsymbol{r}})$ in terms of $f(\hat{\boldsymbol{r}})$ and $g(\hat{\boldsymbol{r}})$. The upshot of the tensor product is that it is normalized in the sense that the CG coefficients constitute unitary matrices converting between product and sum spaces.

The tensor product formulas allow the generation of higher-body-order features from lower-body-order features, starting with elementary features. A single function of the environment $f(\hat{\boldsymbol{r}})$ constitutes a 2-body feature because it describes the correlation between a central atom and its set of neighbors individually. The product of two 2-body features corresponds to a 3-body feature, as the same central atom is now compared against the product set of two neighbors in the environment. To gain intuition for why products of tensors correspond to higher body-order tensors, one can examine, for the sake of example, a cross-correlation function $R(\gamma)$ which probes the correlation between two functions on the sphere offset by an angle of $\gamma$:
\begin{align}
R(\gamma) &= \int f(\hat{{\boldsymbol{x}}}) g(\hat{\boldsymbol{y}}) \delta (\hat{{\boldsymbol{x}}} \cdot \hat{\boldsymbol{y}} - \cos{\gamma}) d^2 \hat{{\boldsymbol{x}}} d^2 \hat{\boldsymbol{y}}\\
&= \sum_{\ell_1,m_1,{\ell_2},{m_2}}   f^{\ell_1}_{m_1} g^{{\ell_2}}_{{m_2}} \int Y^{{\ell_1}}_{{m_1}}(\hat{\boldsymbol{x}})  Y^{{\ell_2}}_{{m_2}}(\hat{\boldsymbol{y}}) \delta (\hat{{\boldsymbol{x}}} \cdot \hat{\boldsymbol{y}} - \cos{\gamma}) d^2 \hat{{\boldsymbol{x}}} d^2 \hat{\boldsymbol{y}}\\
&= \sum_{\ell_1,m_1,{\ell_2},{m_2}}   f^{\ell_1}_{m_1} g^{{\ell_2}}_{{m_2}} \int Y^{{\ell_1}}_{{m_1}}(\hat{\boldsymbol{x}})  Y^{{\ell_2}}_{{m_2}}(\hat{\boldsymbol{y}}) \delta (\hat{{\boldsymbol{x}}} \cdot \hat{\boldsymbol{y}} - \cos{\gamma}) d^2 \hat{{\boldsymbol{x}}} d^2 \hat{\boldsymbol{y}} .
\end{align}
By using orthogonality and completeness relations (in particular, the set of $m=0$ harmonics $Y^\ell_0$ have a completeness relation along $(x,y,z) =(\sin{\theta},0,\cos{\theta})$, the half-circle), one can demonstrate that $R(\gamma)$ can itself be expanded as a tensor along half-circle in terms of only $m=0$ components:
\begin{eqnarray}
R(\gamma) &=& \sum_\ell R^\ell_0 \, Y^\ell_0 (\gamma,0) \\
R^\ell_0 &=& \sum_m M^{\ell \ell 0} f^\ell_m g^{\ell }_{m} .
\end{eqnarray}
This shows that tensor representations of 2-point correlation functions on the sphere, representing angular displacement, can be computed in terms of products of single-neighbor tensors. This principle can be applied inductively to show that  If $\boldsymbol{f}^{\ell_1}$ is an $(n_{f}+1)$-body tensor and and $\boldsymbol{g}^{\ell_2}$ is an $(n_{g}+1)$-body tensor, respectively, then $\boldsymbol{h}^{\ell_3}$ is an $(n_{f}+n_{g}+1)$-body-order tensor.

Finally, in considering a tensor $\boldsymbol{f}^\ell$, if $\ell=0$, then $f$ is a scalar with a single, rotationally invariant component, $f^0_0$, which makes $f^0_0$ a suitable feature for regression to scalar quantity, e.g.\ atomic charge or potential energy.  A variety of methods such as ACE and others apply this scheme systematically to a family of functions on the sphere built from differing density projection functions to yield a family of features suitable for potential energy regression, or, in the case of $\ell>0$ features, equivariant regression. An important aspect of this expansion is that it yields a combinatoric number of potential high-order features. 

Enumerating all possible invariants according to the spherical tensor approach is somewhat complex: one must track not only the underlying functions and their tensor ranks, but also the order in which tensor product terms are generated and the order of those intermediate tensor products. Not all values of intermediate orders $\ell$ are allowed; they must fall between the bounds given by the input tensors. Finally, the whole apparatus must be combined with the Clebsh-Gordon coefficients to yield the resulting tensors.

\subsection{ACE Basis Reduction with the Langbein Algorithm}
\begin{figure*}[htb]
    \centering
    \includegraphics[width=6.75in]{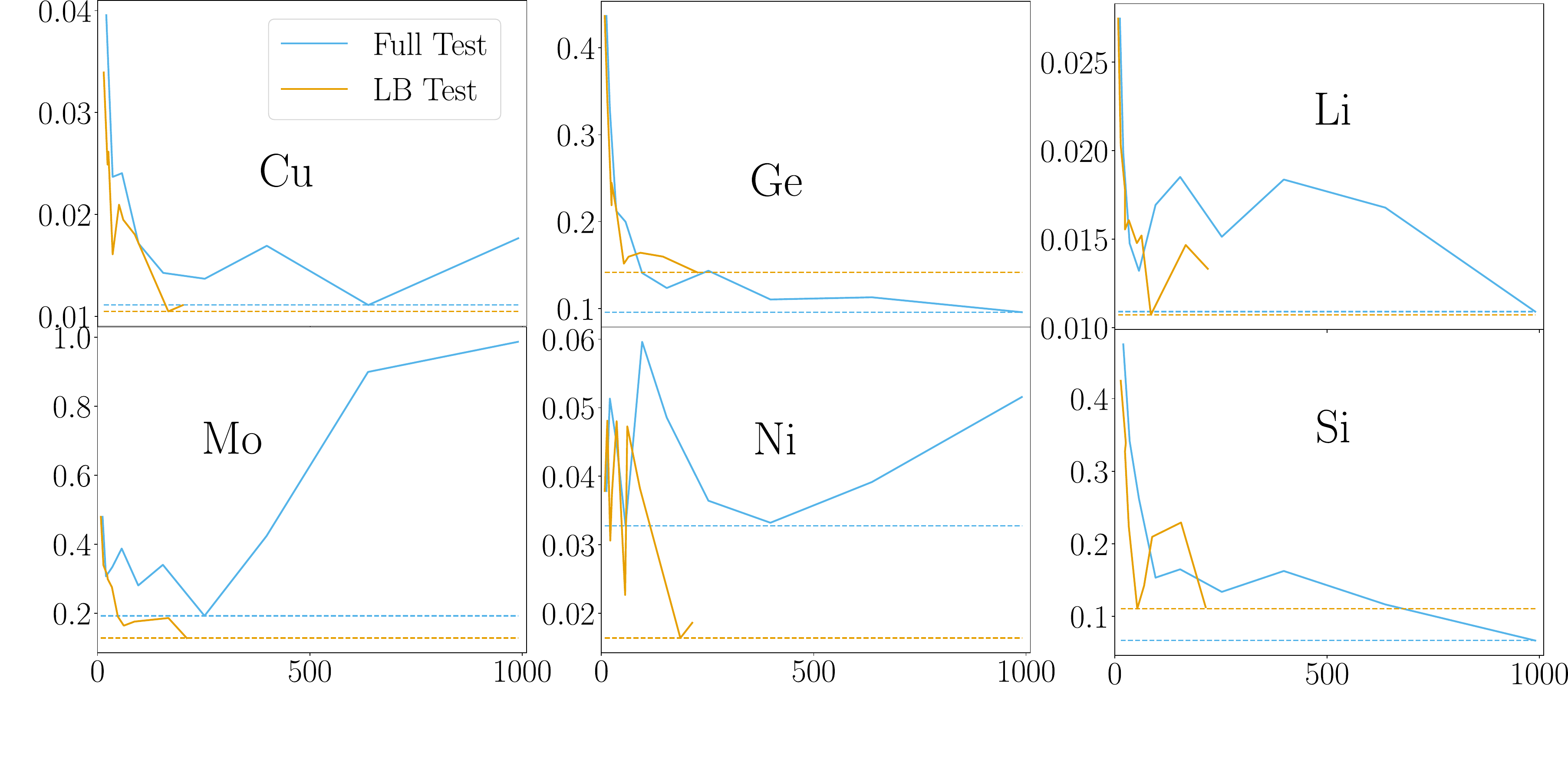}
    \caption{
    The change in test set energy error with the size of the basis set for the six materials present in Ref.~\citenum{Zuo2020}. A comparison of the full basis set and the basis set reduced with the Langbein algorithm is shown. The dashed lines show the lowest error achievable for the full basis set and Langbein subset.}
    \label{fig:ACE_Energy_Langbein}
\end{figure*}

The energy error with basis set size for the ACE full model and the Langbein subset is shown in Fig.~\ref{fig:ACE_Energy_Langbein}. There are significantly fewer energy values compared to force components, leading to increased fluctuations for theses results. 

\subsection{Training Details for Methane Configurations Dataset}\label{SI:methane}
The same model hyperparameters were used for each network type and training set size. Each model has one interaction layer, three atom layers with 32 features each, and 20 sensitivity functions. Other hyperparameters such as learning rate, batch size, and patience were also kept constant. The hyperparameters were selected with the aim of creating models that could be quickly trained, so that the 440 total models could be trained while allowing enough expressivity that the models could learn well. The models were trained to match both forces and energies associated with the configurations.

\subsection{Hyperparameter Search Details for QM7 Models}
The hyperparameter search for the QM7 models was performed using a Tree-structured Parzen Estimator~\cite{Bergstra2011AlgorithmsFH}, a Bayesian optimization method frequently employed for hyperparameter tuning. The search was implemented with Ray Tune~\cite{raytune} and Hyperopt~\cite{hyperopt}. There were 200 trials conducted for each network type, with the results for each trial being an average from 10 trained models on a specific set of hyperparameters. The search space consisted of seven hyperparameters, the best found combinations of which are listed in Table~\ref{tab:qm7_hyperparams}. Each model was trained using a random 70\% of the QM7 dataset. An additional 10\% was used to measure early-stopping criteria for model training and to quantify model performance for the hyperparameter search. The remaining 20\% in each case was not used during the hyperparameter search. The results in Table~\ref{tab:qm7_results} are calculated from 10 models using the best hyperparameters found for each network type, again using a random $70+10\%$ of the QM7 data. The MAE and RMSE values reported in the table are on the final 20\% of the data. An example code for performing this hyperparameter search can be found in the the open-source \texttt{hippynn}~\cite{hippynn-repo} repository, which implements the HIP-NN family of models. 

\begin{table}
\begin{tabular}{|c|c|c|c|c|}
\hline
\multicolumn{5}{|c|}{\textbf{Architecture}} \\
\hline
\textbf{Model}              & HIP-NN       & HIP-NN-TS    & HIP-NN-TS & HIP-HOP-NN      \\
\hline
$\mathbf{n}$                & 1            & 2            & 2         & 4               \\
\hline
\boldmath$\ell$\unboldmath  & 0            & 1            & 2         & 3               \\
\hline
\multicolumn{5}{|c|}{\textbf{Results}} \\
\hline
$\mathbf{n_\mathrm{interaction}}$      & 2              & 1                 & 1                 & 1                 \\
\hline
$\mathbf{n_\mathrm{on-site}}$          & 2              & 1                 & 4                 & 5                 \\
\hline
$\mathbf{n_\mathrm{feature}}$          & 96             & 122               & 122               & 165               \\
\hline
$\mathbf{n_\mathrm{sensitivity}}$      & 28             & 13                & 19                & 20                \\
\hline
\textbf{learning rate}                 & $1.91\times 10^{-3}$  & $3.10\times 10^{-4}$  & $7.11\times 10^{-4}$  & $5.65\times 10^{-4}$  \\
\hline
\textbf{batch size}                    & 25             & 21                & 27                & 100               \\
\hline
\textbf{patience}                      & 185            & 99                & 166               & 183                  \\
\hline

\end{tabular}
\centering
\begin{caption}{Best hyperparameter set found for each model for the QM7 dataset}\label{tab:qm7_hyperparams}
\end{caption}
\end{table}

\begin{figure}[htbp]
  \centering
  \begin{subfigure}[b]{0.47\textwidth}
    \centering
    \includegraphics[width=\textwidth]{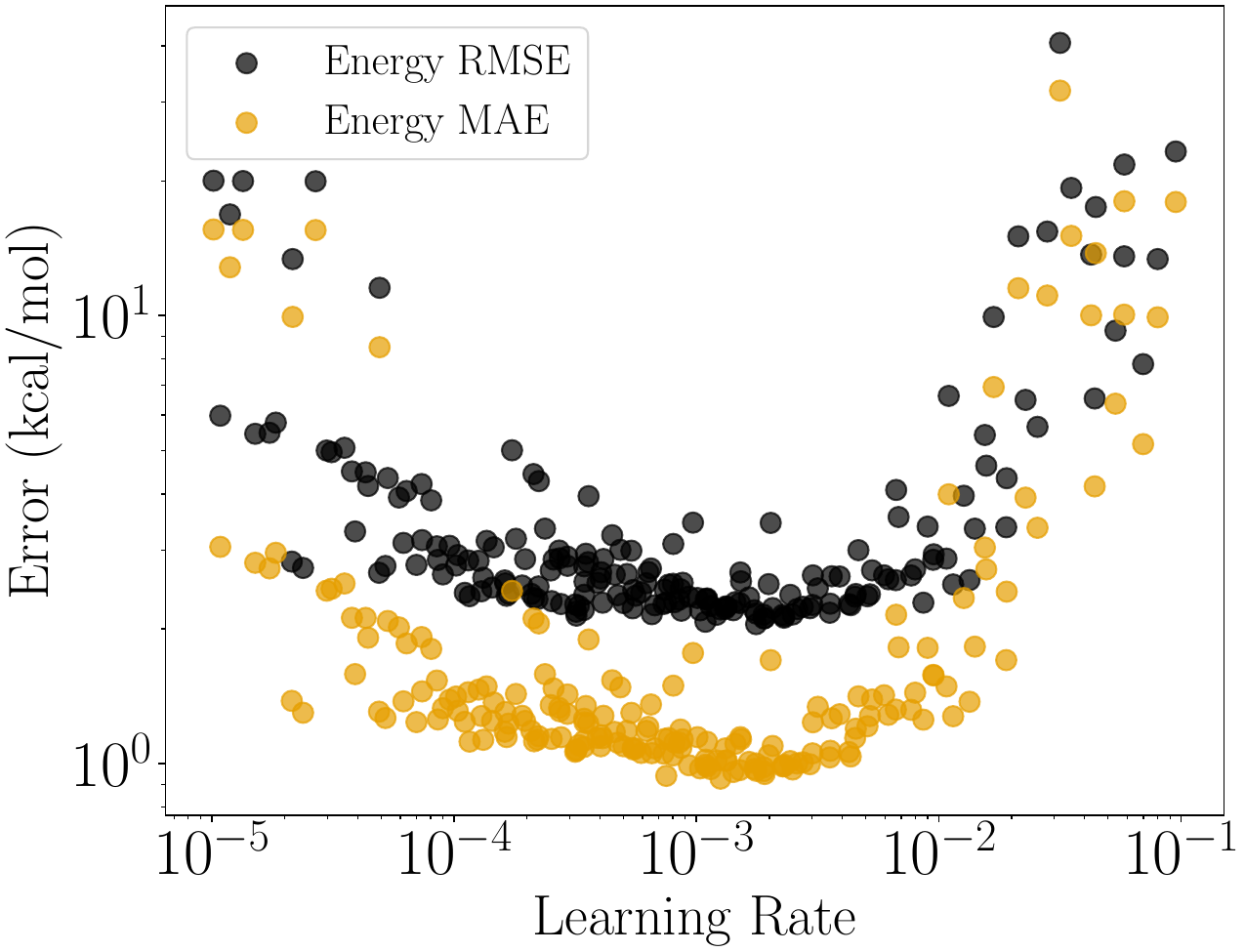}
    \caption{HIP-NN}
  \end{subfigure}
  \hfill
  \begin{subfigure}[b]{0.47\textwidth}
    \centering
    \includegraphics[width=\textwidth]{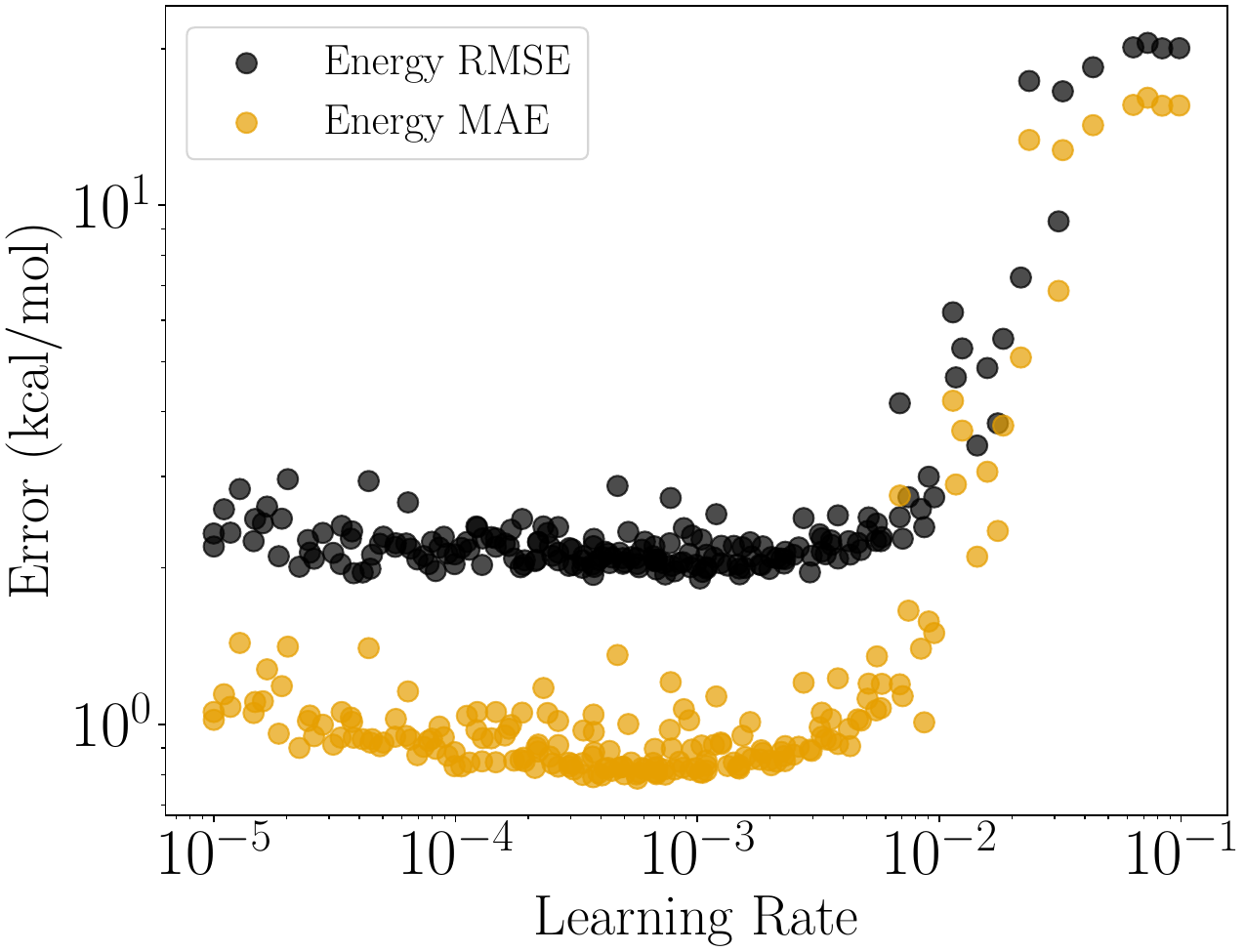}
    \caption{HIP-HOP-NN}
  \end{subfigure}
  \caption{Error metrics on validation set for each set of hyperparameters considered across the search, organized by learning rate}
  \label{fig:hyperparam_results_si}
\end{figure}

We observed that the learning rate is the most significant hyperparameter as well as the hyperparameter whose optimal values varied most significantly between architectures. Figure~\ref{fig:hyperparam_results_si} shows the mean validation scores for all 200 sets of hyperparameters probed organized by learning rate for the HIP-NN and HIP-HOP-NN models. The optimal learning rate for HIP-HOP-NN is about a factor of two smaller than the optimal learning rate for HIP-NN.  

\subsection{Hyperparameters for QM9 and ANI-1x HIP-HOP-NN Models}

The HIP-HOP-NN hyperparameters used for the QM9 dataset and ANI-1x dataset are shown in Table \ref{tab:hyperparams}. The ANI-1x model is also used for the speed tests in this work. For more detail about the hyperparameters set up see Ref.~\citenum{Chigaev2023}.

\begin{table}
\begin{tabular}{|r|c|c|}
\hline
\textbf{Model} & \textbf{QM9} & \textbf{ANI-1x} \\ \hline
$\mathbf{n_\mathrm{interaction}}$ & 2 & 2 \\ \hline
$\mathbf{n_\mathrm{on-site}}$  & 6 & 5 \\ \hline
$\mathbf{n_\mathrm{feature}}$ & 256 & 256 \\ \hline
$\mathbf{n_\mathrm{sensitivity}}$ & 20 & 20 \\ \hline
Cutoff Distance & 6.5 & 6.5 \\ \hline
Lower Cutoff & 0.85 & 0.75 \\ \hline
Moment Tensor Order & 3 & 3 \\ \hline
\end{tabular}
\caption{The HIP-HOP-NN hyperparameters used for the QM9 and ANI-1x benchmark problems. }
\label{tab:hyperparams}
\end{table}

\subsection{Force Training Schedule for QM9}
The QM9 dataset contains equilibrium structure with zero values for all force components. These forces can be added to the training protocol for MLIPs to improve the recreation of energies. The force weighting in the loss function is set to 1 for the first 10 epochs, 2 for 50 epochs and then 0.1 for the remaining training. 

\clearpage
\bibliography{references}

\clearpage

\end{document}